\def\extendedbuild{1}
\documentclass{article}
\usepackage[T1]{fontenc}
\usepackage{spconf,amsmath,amsfonts,graphicx,hyperref}
\usepackage{booktabs}
\usepackage{array}
\usepackage[caption=false,font=footnotesize]{subfig}
\usepackage{url}
\hypersetup{hidelinks}
\newif\ifappendix
\ifdefined\extendedbuild
  \appendixtrue
\else
  \appendixfalse
\fi
\ifappendix
  \usepackage{multibib}
  \newcites{app}{Appendix References}
\fi

\makeatletter
\def\thebibliography#1{\section{References}\list
 {[\arabic{enumi}]}{\settowidth\labelwidth{[#1]}\leftmargin\labelwidth
 \advance\leftmargin\labelsep\itemsep 1pt\parsep 0pt
 \usecounter{enumi}}
 \def\newblock{\hskip .11em plus .33em minus .07em}
 \sloppy\clubpenalty4000\widowpenalty4000
 \sfcode`\.=1000\relax}
\makeatother
\graphicspath{{figs/}}

\usepackage{eso-pic}
\AddToShipoutPictureFG*{%
  \put(0,34){\makebox[\paperwidth]{\parbox{6.4in}{\centering\footnotesize
    This work has been submitted to the IEEE ICASSP for possible publication.
    Copyright may be transferred without notice, after which this version
    may no longer be accessible.}}}}
\title{REDIMNET2+: MULTI-CORPUS DATA SCALING FOR ROBUST SPEAKER VERIFICATION}

\name{Kirill Borodin$^{1,2}$, Vasilii Kudryavtsev$^{1,2}$, Maxim Maslov$^{1}$, Grach Mkrtchian$^{2,3}$}
\address{$^{1}$lab260, Yerevan, Armenia \quad $^{2}$BitmanagerAI, Dubai, UAE \quad $^{3}$MTUCI, Moscow, Russia \\ kborodin.research@gmail.com}

\begin{document}
\ninept
\maketitle

\begin{abstract}
Automatic speaker verification must remain reliable across devices, rooms, and compression pipelines. We present ReDimNet2+, which scales training of the compact ReDimNet2 backbone across seven public corpora (63,934 speakers, about 8,675 hours). Analysis of a VoxBlink2 subset reveals a shift in predicted spectral coloration, motivating codec and waveform augmentation alongside this multi-corpus training, large-margin fine-tuning (LMFT), and graph-based retrieval reranking. With random 4-second evaluation windows for all models, ReDimNet2+ LMFT reduces pooled VoxCeleb1 EER from 2.42\% to 0.82\% and a 26-condition robustness stress-test EER from 7.21\% to 1.99\%. Under this shared local protocol, it reaches 0.35\% EER on VoxCeleb1-O versus 0.787\% for the best evaluated WeSpeaker checkpoint. On a VoxBlink2 retrieval subset, reranking improves the final model's Pr@$k$ from 0.7413 to 0.7687.
\end{abstract}


\section{Introduction}
\label{sec:intro}
Automatic speaker verification (ASV) maps a speech segment to a representation that should retain speaker identity while discarding nuisance factors such as microphone response, room acoustics, noise, compression, language, and duration. Modern systems reach low error rates on clean benchmarks such as VoxCeleb1 \cite{vc1,vc2}, yet degrade under unseen channels, aggressive compression, short fragments, or retrieval-style open-set evaluation. Large public corpora such as VoxBlink2 \cite{vb2}, 3D-Speaker \cite{3d_speaker}, CN-Celeb \cite{cnceleb,cnceleb2}, TidyVoice \cite{tidy}, and KeSpeech \cite{kespeech} add speaker and condition diversity, but data alone does not guarantee robustness if the target degradation mechanisms are absent from training.

Robustness has been pursued through noise, reverberation, and metric learning \cite{10096848,Mao2020ShortTimeSV,musan,rir}, feature-level regularization \cite{park19e_interspeech,Yun2019CutMix}, margin objectives \cite{sphereface,li2026languageinvariantmultilingualspeakerverification}, and large self-supervised encoders \cite{wavlm,w2v2bert,li2026enhancingspeakerverificationw2vbert} whose cost often requires distillation. Open-set identification additionally depends on nearest-neighbor ranking, which reciprocal-neighbor reranking \cite{Zhong2017reranking} and hubness correction \cite{suzuki2013centering} address in other domains.

This paper studies robust ASV around ReDimNet2, a compact backbone based on time-pooled dimension reshaping that is competitive with much larger self-supervised front ends \cite{redimnet2}. Its public checkpoint, however, is not adapted to the channel and codec mismatch observed in our target setting. We keep the backbone architecture fixed and ask how far robustness can be improved through data, augmentation, optimization schedule, and scoring alone.\footnote{\urlstyle{same}Code: \url{https://github.com/lab260ru/redimnet2-plus}. Pretrained weights: \url{https://huggingface.co/lab260/redimnet2-plus}.} Our contributions are:
\begin{itemize}\setlength\itemsep{0pt}
\item a data-driven analysis of a VoxBlink2 subset identifying a shift in predicted coloration that motivates channel and codec augmentation (Sec.~\ref{sec:data});
\item an efficient pipeline with random-window decoding, explicit codec simulation, waveform and feature augmentation, and a staged multi-corpus recipe with large-margin fine-tuning (LMFT) that reduces pooled VoxCeleb1 EER from 2.42\% to 0.824\%;
\item a lightweight graph-based reranking stage for open-set retrieval exploiting local neighborhood consistency and hubness penalties;
\item a reproducible local comparison against open WeSpeaker checkpoints on VoxCeleb1-O/E/H, using a shared 4-second evaluation window and distinguishing locally reproduced results from published full-utterance ReDimNet2 results.
\end{itemize}

\section{Data and Domain Mismatch}
\label{sec:data}
\textbf{Training corpora.} The final training pool combines seven corpora with 63,934 speakers, about 4.6M utterances, and about 8,675 hours in total: VoxBlink2 \cite{vb2} and VoxCeleb2 \cite{vc2} provide large in-the-wild YouTube speech; 3D-Speaker \cite{3d_speaker} adds device, distance, and dialect variation; CN-Celeb and CN-Celeb2 \cite{cnceleb,cnceleb2} add multi-genre Chinese web speech; TidyVoice \cite{tidy} contributes read speech in 62 languages derived from Common Voice; and KeSpeech \cite{kespeech} contributes Mandarin and subdialect read speech. Checkpoints are selected on the VoxCeleb1 development set. VoxCeleb1 is used for verification and retrieval evaluation.\ifappendix\ Per-corpus statistics are given in appendix Table~\ref{tab:data}.\fi

\textbf{VoxBlink2 subset analysis.} The exploratory analysis used a VoxBlink2 subset: 673,277 training FLAC files from 11,053 speakers ($\approx$1,458 h) and an evaluation subset of 134,697 files ($\approx$345 h) used for our speaker-retrieval evaluation (Pr@$k$), all 16~kHz mono. Speaker counts are strongly imbalanced (69 speakers with one recording, 4,870 with 76--100). Training utterances are shorter: 22.7\% are below 3~s, whereas the evaluation subset contains more 5--20~s clips\ifappendix\ (appendix Table~\ref{tab:duration})\fi. Early stages therefore use 2--3~s random crops, and LMFT uses 6~s windows.

\textbf{Coloration and byte-rate shift.} Median FLAC byte rate drops from 21,040 bytes/s in train to 14,021 bytes/s in the evaluation subset, and the fifth percentile from 16,283 to 8,989 bytes/s. We treat byte rate as a heuristic indicator of signal compressibility, not a direct bandwidth measurement: it also depends on encoding settings, silence, and acoustic content. NISQA coloration (COL) predictions \cite{nisqa} provide complementary evidence: 6.2\% of training files have COL below 2.0, versus 41.4\% of evaluation files.\ifappendix\ Detailed statistics appear in appendix Table~\ref{tab:bytes_col}.\fi\ For 884 speakers with at least five recordings, the median within-speaker standard deviation of bytes/s is 1,753; 41\% exceed 2,000 bytes/s. These observations are consistent with heterogeneous recording conditions, although byte-rate variation alone does not establish channel distortion. The analysis motivates testing codec simulation and filtering alongside noise and reverberation augmentation.

\section{ReDimNet2+}
\label{sec:method}

\subsection{Training pipeline and augmentation}
\label{ssec:pipeline}
\begin{figure}[!t]
\centering
\includegraphics[width=\linewidth]{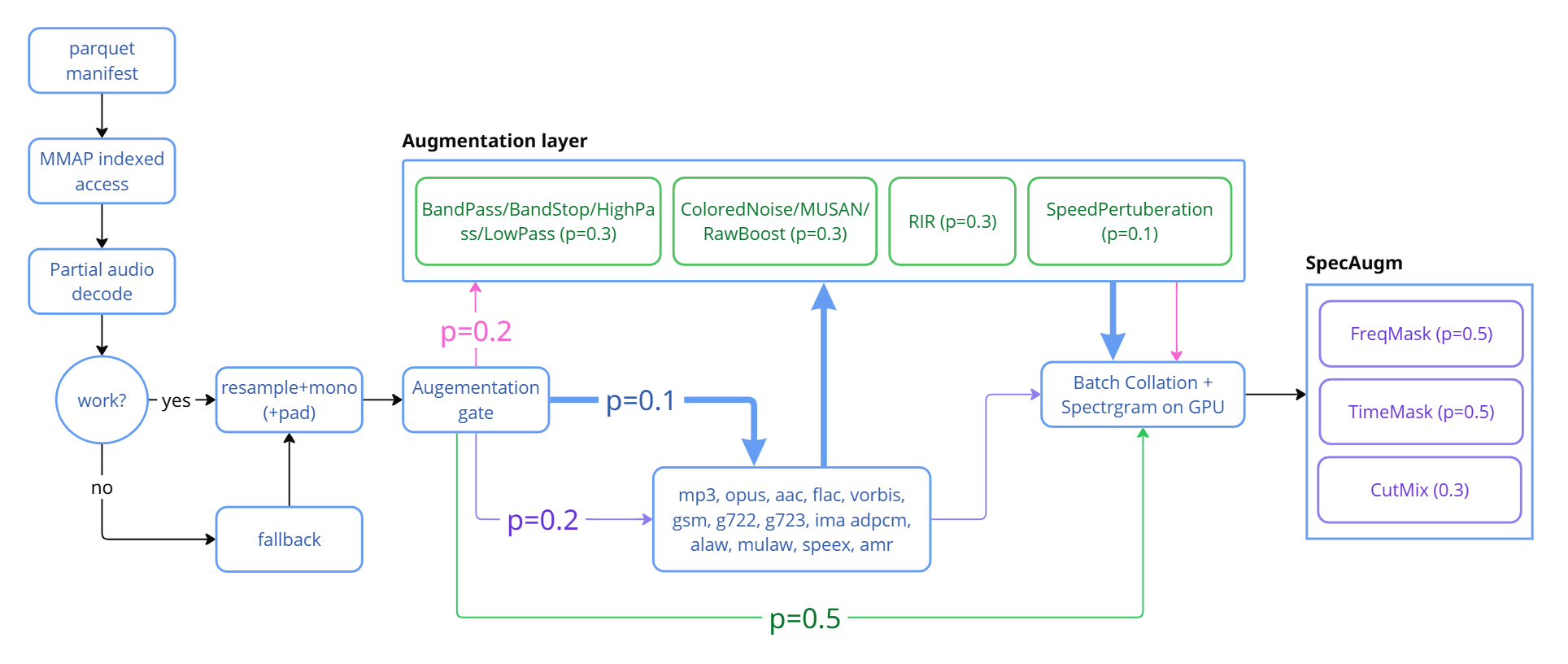}
\caption{Training-time data pipeline. Audio is indexed through Parquet metadata, decoded as a random window, augmented at waveform and codec level, batched as waveforms, and converted to spectrogram features on the GPU.}
\label{fig:pipeline}
\end{figure}

\begin{figure*}[!t]
\centering
\subfloat[clean]{\includegraphics[width=0.16\textwidth]{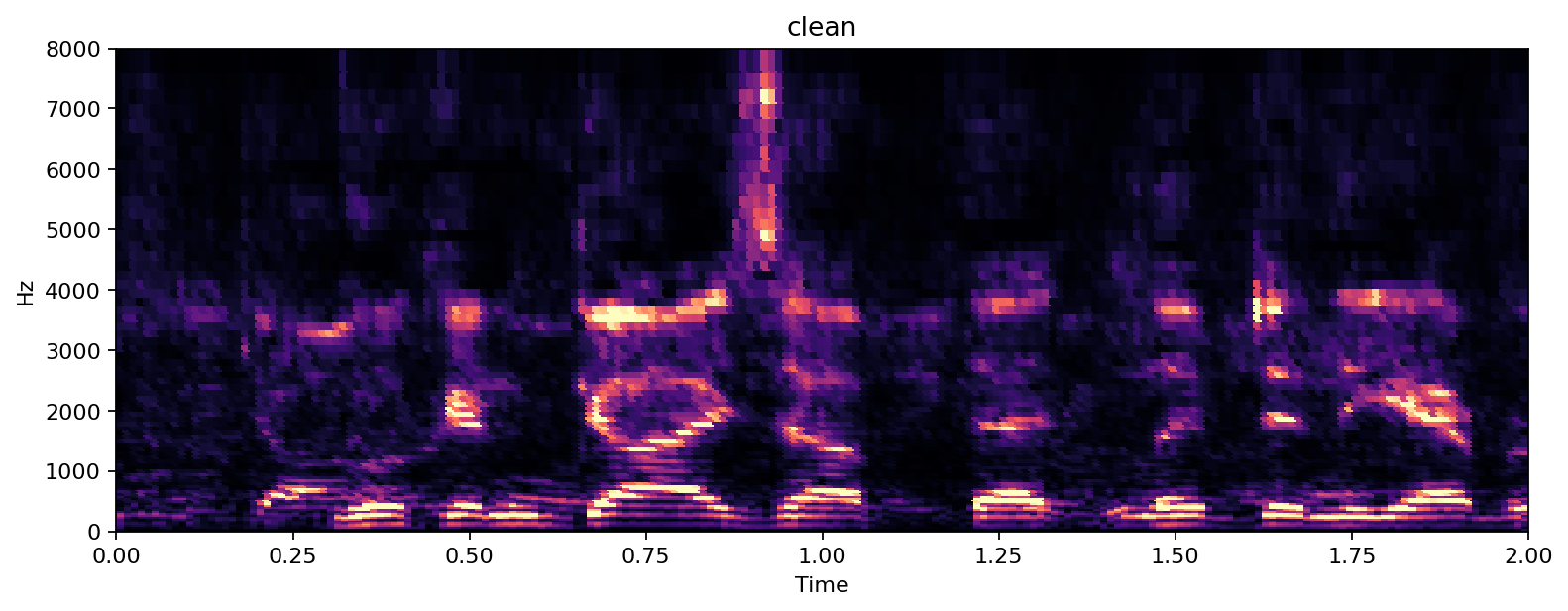}}
\hfil
\subfloat[AAC 16~kHz]{\includegraphics[width=0.16\textwidth]{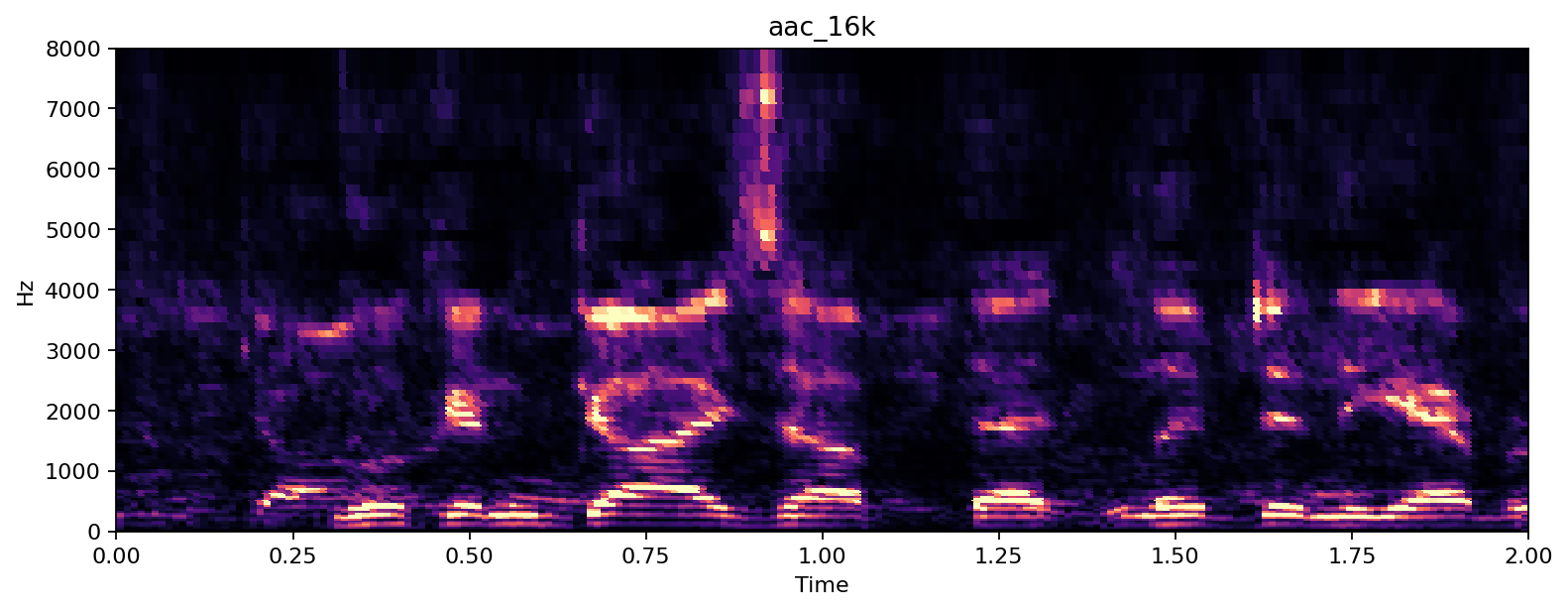}}
\hfil
\subfloat[AMR-NB]{\includegraphics[width=0.16\textwidth]{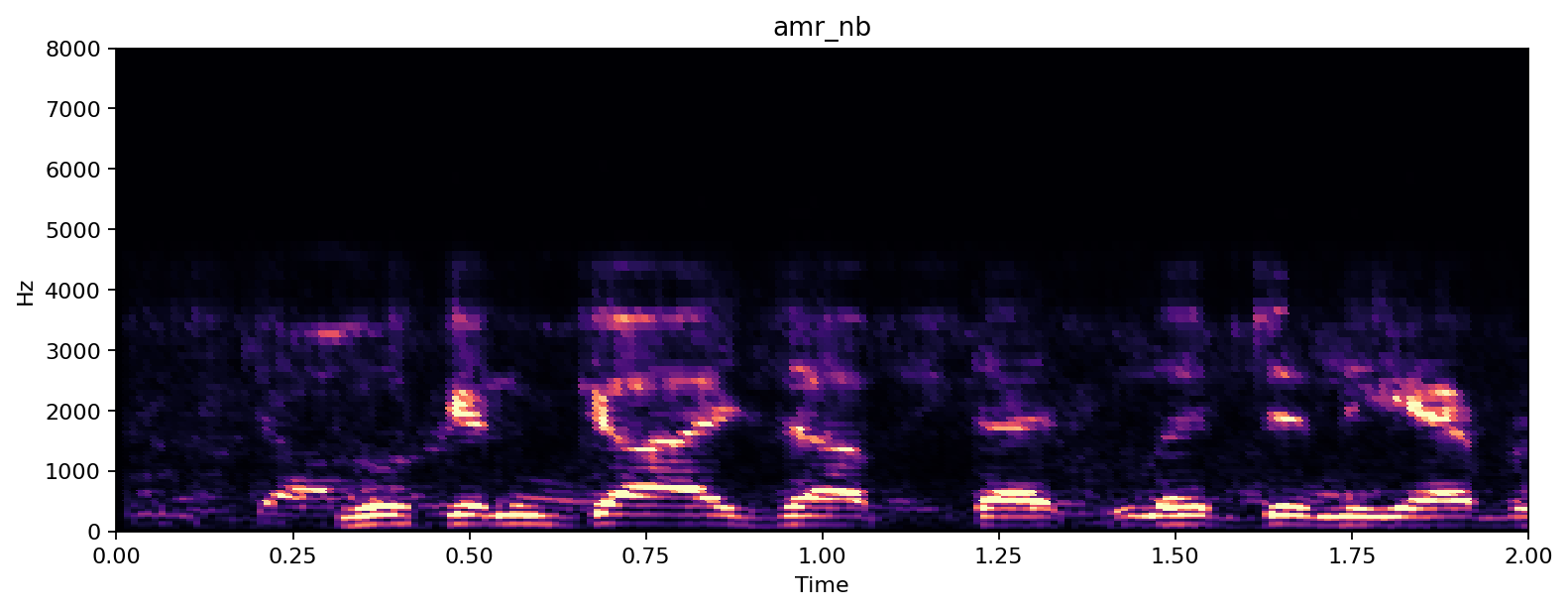}}
\hfil
\subfloat[G.722]{\includegraphics[width=0.16\textwidth]{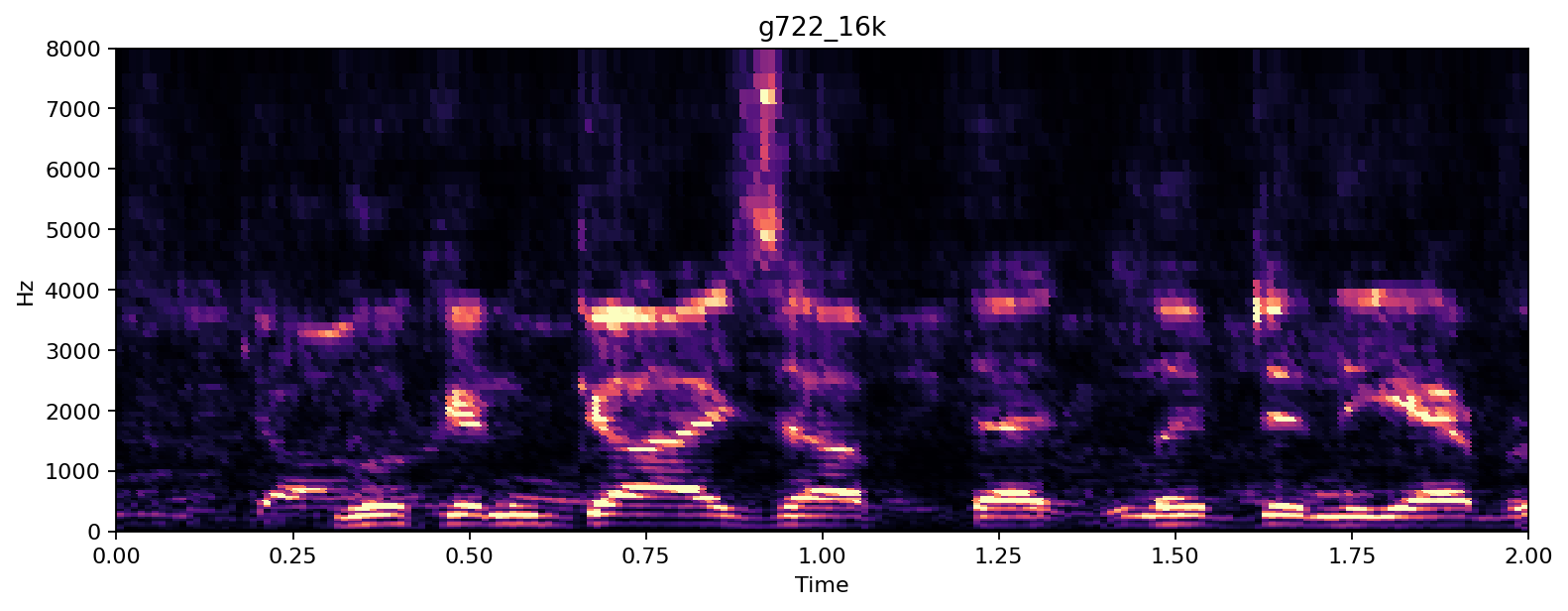}}
\hfil
\subfloat[$\mu$-law 8~kHz]{\includegraphics[width=0.16\textwidth]{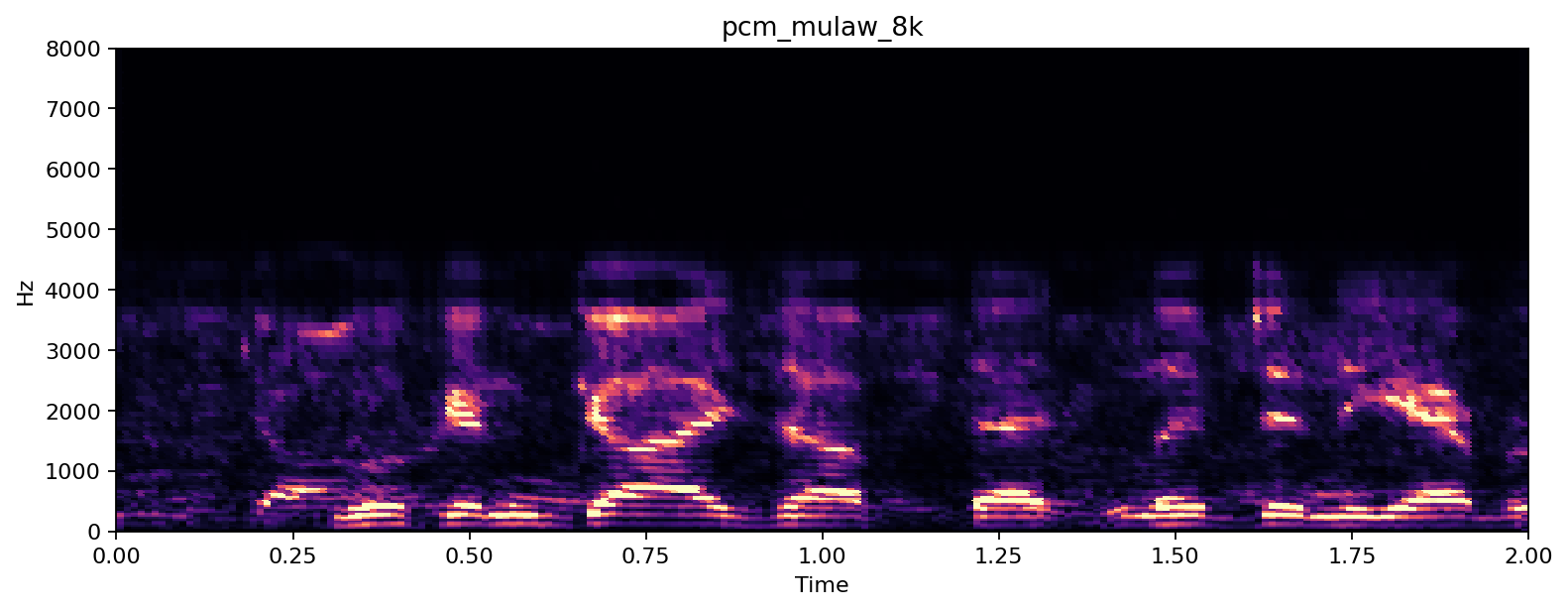}}
\hfil
\subfloat[Speex 16~kHz]{\includegraphics[width=0.16\textwidth]{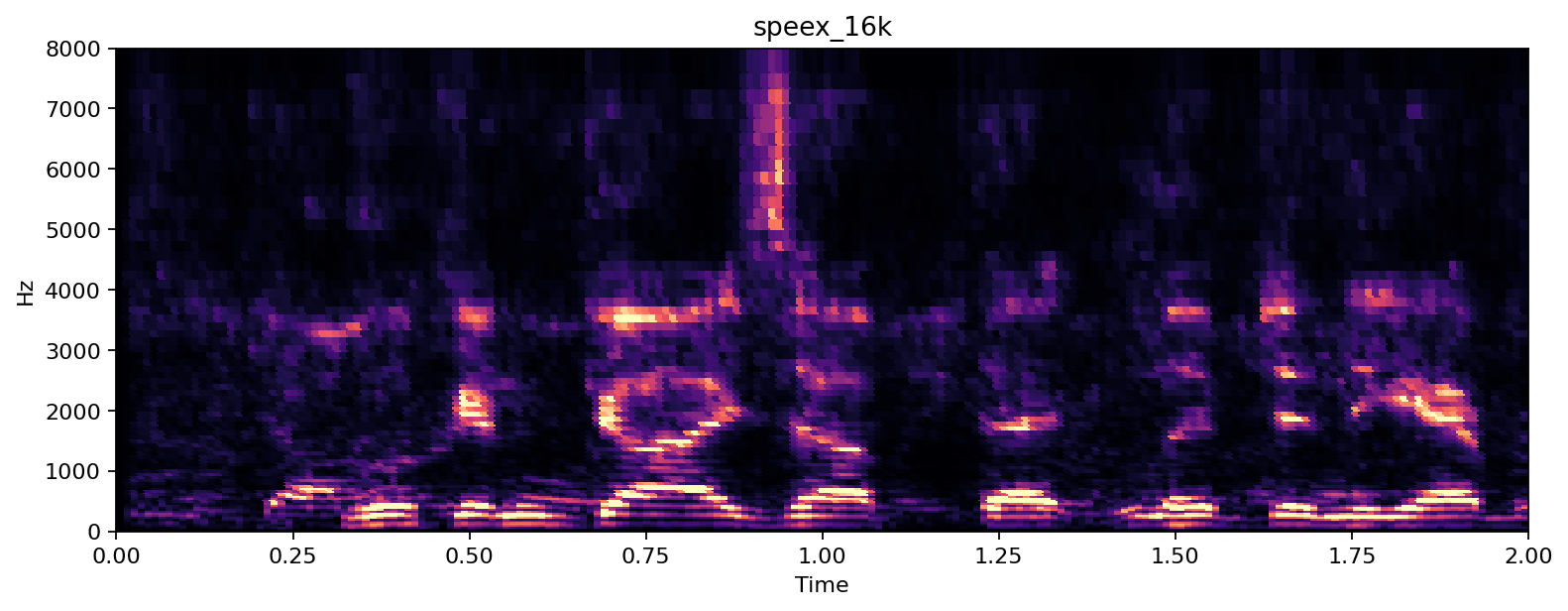}}
\caption{Examples of codec transformations used to model spectral coloration and bandwidth limitation: the same utterance after different codec presets.}
\label{fig:codec_aug}
\end{figure*}

Fig.~\ref{fig:pipeline} shows the pipeline. Training reads a Parquet manifest and decodes only a random fixed-length window of each file, which is 2.6$\times$ faster than full decoding followed by cropping; on a decode failure the loader falls back to a neighboring example so that isolated corrupt files do not interrupt multi-day runs on crawled corpora. All audio is mono 16~kHz, and short segments are repeat-padded, which preserves speech content and avoids artificial silence tails at the long LMFT window. Speed perturbation is restricted to ten fixed resampling factors in $[0.9,1.1]$ for kernel reuse, codec augmentation calls FFmpeg directly rather than through a Python wrapper, and spectrogram computation is moved from the CPU collator to the GPU training step, which improved throughput by 1.53$\times$ with four loader workers\ifappendix\ (appendix Table~\ref{tab:pipeline_speed})\fi.

Each example is assigned to one of four modes with probabilities 0.5/0.2/0.2/0.1: keep, codec only, waveform augmentation only, or codec plus waveform. The waveform branch contains filters (band-pass, band-stop, high-pass, low-pass), additive noise (colored noise, MUSAN \cite{musan}, RawBoost-style corruption), simulated room impulse responses \cite{rir}, and speed perturbation, each applied with probability 0.3/0.3/0.3/0.1. The codec branch simulates compression and telephony channels with MP3, Opus, AAC, FLAC, Vorbis, G.723.1, IMA ADPCM, G.722, A-law, $\mu$-law, Speex, and AMR-NB presets, several at both 8 and 16~kHz (Fig.~\ref{fig:codec_aug}); it is directly motivated by the coloration shift of Sec.~\ref{sec:data}. SpecAugment and CutMix \cite{park19e_interspeech,Yun2019CutMix} are applied after GPU spectrogram extraction with probability 0.2. FLAC is lossless and is not itself a source of codec distortion.\ifappendix\ The full policy and waveform-level examples are given in appendix Table~\ref{tab:aug_policy} and Fig.~\ref{fig:waveform_aug}.\fi

\subsection{Staged fine-tuning and LMFT}
\label{ssec:training}
We initialize ReDimNet2 from its public checkpoint \cite{redimnet2}. Training uses a SphereFace2 head \cite{sphereface} with one-versus-all binary classification objectives for normalized speaker embeddings and class weights. We follow the cited formulation and vary the margin during adaptation. At evaluation, L2-normalized embeddings are scored by cosine similarity.

Every adaptation run in Table~\ref{tab:main_results} starts independently from the public checkpoint, except the final ReDimNet2+ LMFT run. ReDimNet2+ pretrained adapts that checkpoint on all seven training corpora; the final LMFT run continues from this adapted model. The \emph{LMFT only} control applies LMFT directly to the public checkpoint. Checkpoints are selected on the VoxCeleb1 development set. Across the independent adaptation runs, training windows increase from 32,200 to 48,300 samples and the data pool broadens. LMFT uses a fixed margin of 0.3 and 96,000-sample (6~s) training windows. Because learning rates and training duration also vary, these runs characterize the combined recipe rather than isolate individual factors. All runs use Accelerate with FSDP on six GPUs, BFloat16, AdamW, gradient clipping at 20, and seed 42.\ifappendix\ Per-run hyperparameters are listed in appendix Table~\ref{tab:hparams}.\fi

\subsection{Retrieval reranking}
\label{ssec:rerank}
Let $k$ denote the retrieval evaluation cutoff and $K$ the graph neighborhood size. A model with low EER may still suffer from hubness or local ranking errors in a large gallery: a few ``universal'' embeddings enter the top-$k$ lists of many queries, and recordings of one speaker under different channels form sub-clusters that cosine distance ranks inconsistently. We therefore apply two lightweight reranking stages on the L2-normalized embeddings; both operate on cached embeddings and neighbor lists, need no retraining, and leave verification scores untouched.

\textbf{Mean-chain reranking.} For query $e_i$, an initial candidate pool $P(i)$ of size $M>K$ is selected by cosine similarity. Instead of taking the $K$ nearest candidates independently, the stage follows a locally consistent chain. Starting from the probe $p_i^{(0)}=e_i$, step $t$ picks the unused candidate closest to the current probe,
\begin{equation}
j_t=\arg\max_{j\in P(i)\setminus\{j_1,\ldots,j_{t-1}\}} e_j^{T}p_i^{(t-1)},
\end{equation}
and moves the probe to the normalized mean of the query and the selected neighbor,
\begin{equation}
p_i^{(t)}=(e_i+e_{j_t})/\|e_i+e_{j_t}\|_2 .
\end{equation}
Anchoring the probe to $e_i$ prevents drift away from the query, while the neighbor term lets the ordering follow the local structure of the speaker cluster. After $K$ selections, the chain order forms the neighbor list $N(i)$ passed to the next stage. The original query remains in every probe update; the update averages it with the most recently selected neighbor, rather than with all previously selected neighbors.

\textbf{Graph reranking.} The second stage uses neighbor-list structure without speaker labels. It builds a $K$-NN graph from the neighbor lists and rescores, for each query, the candidates among its neighbors and their neighbors,
\begin{equation}
P_{\mathrm{graph}}(i)=N(i)\cup\bigcup_{u\in N(i)}N(u),\qquad i\notin P_{\mathrm{graph}}(i),
\end{equation}
and each candidate receives
\begin{equation}
S(i,j)=w_rR(i,j)+w_qQ(i,j)+w_cC(i,j)-w_hH(j),
\end{equation}
where $R$ rewards a high rank of $j$ in $N(i)$, $Q$ rewards reciprocal support ($i\in N(j)$) in the spirit of $k$-reciprocal reranking \cite{Zhong2017reranking}, and $C$ rewards common-neighbor support, i.e. candidates whose neighborhood overlaps with that of $i$. The hubness term
\begin{equation}
H(j)=\max\!\left(0,\log\frac{\mathrm{deg}_{in}(j)}{K}\right)
\end{equation}
penalizes candidates whose in-degree exceeds $K$, the average in-degree of a $K$-NN graph, a degree-based counterpart of hubness correction \cite{suzuki2013centering}. Candidates are reordered by $S(i,j)$, the top-$K$ become the new $N(i)$, and the graph is rebuilt; the final recipe uses $w_r{=}1.0$, $w_q{=}1.5$, $w_c{=}2.0$, $w_h{=}0.3$ and three such iterations.

\section{Experiments}
\label{sec:exp}
\begin{table*}[!t]
\caption{Main ReDimNet2+ development path on VoxCeleb1 (4-second evaluation windows; EER, \%). EER$_p$ pools O/E/H scores; EER$_{ph}$ is the 26-condition codec/waveform stress test; Pr@$k$ is the VoxBlink2-subset retrieval metric before reranking.\label{tab:main_results}}
\centering
\small
\begin{tabular}{lrrrrrr}
\toprule
System & EER$_o$ & EER$_e$ & EER$_h$ & EER$_p$ & EER$_{ph}$ & Pr@$k$ \\
\midrule
ReDimNet2 baseline & 1.601 & 1.725 & 3.039 & 2.420 & 7.214 & 0.5978 \\
ReDimNet2 (FT) & 2.414 & 2.337 & 4.260 & 3.408 & 5.599 & 0.4890 \\
ReDimNet2 (cosine margin schedule) & 2.143 & 2.236 & 4.122 & 3.271 & 5.502 & 0.4813 \\
ReDimNet2 (VC2, 6 epochs) & 1.734 & 1.645 & 3.010 & 2.404 & 4.624 & 0.6770 \\
ReDimNet2 (multi-domain, 6 epochs) & 1.095 & 1.067 & 1.997 & 1.587 & 3.702 & 0.6906 \\
ReDimNet2 (TidyVoice added) & 0.989 & 1.022 & 1.934 & 1.537 & 3.628 & 0.6936 \\
ReDimNet2 (KeSpeech added, 6 epochs) & 0.936 & 1.034 & 1.958 & 1.559 & 3.522 & 0.6986 \\
ReDimNet2 (multi-domain, 10 epochs) & 0.872 & 0.958 & 1.833 & 1.458 & 3.303 & 0.7005 \\
ReDimNet2+ pretrained & 0.792 & 0.956 & 1.811 & 1.424 & 3.229 & 0.7024 \\
ReDimNet2 (LMFT only) & 0.436 & 0.589 & 1.118 & 0.885 & 2.281 & 0.7221 \\
ReDimNet2+ LMFT & \textbf{0.351} & \textbf{0.523} & \textbf{1.055} & \textbf{0.824} & \textbf{1.991} & \textbf{0.7413} \\
\bottomrule
\end{tabular}
\end{table*}

\subsection{Metrics and protocol}
\label{ssec:protocol}
All models are evaluated using randomly selected 4-second input windows and cosine similarity of L2-normalized embeddings. Checkpoints are selected on the VoxCeleb1 development set. We report EER on VoxCeleb1-O, -E, and -H, and pooled EER$_p$ over the union of the three score lists. As a robustness stress test we use EER$_{ph}$: 26 degraded variants of the VoxCeleb1-O utterances are generated with the waveform and codec transformations of Sec.~\ref{ssec:pipeline}, and a single EER is computed over the concatenated scores. By construction the metric is severe: it asks whether speaker separation survives many channel transformations observed simultaneously through one verification protocol. For retrieval, Pr@$k$ is the fraction of same-speaker items among the first $k$ results, averaged over queries. We evaluate two datasets: VoxCeleb1 (Pr@1/10/45, reported in percent) and a VoxBlink2 subset (Pr@$k$, reported as a fraction). The latter is our subset retrieval protocol, distinct from the published VoxBlink2 open-set identification benchmark.

For the external comparison, the WeSpeaker \cite{wespeaker_toolkit} ResNet34-LM, ECAPA512-LM, and CAM++ ONNX checkpoints \cite{hf_wespeaker_resnet34_lm,hf_wespeaker_ecapa512_lm,hf_wespeaker_campplus} and the public ReDimNet2 checkpoint were evaluated on the same local VoxCeleb1 trial files (37,611 / 579,818 / 550,894 pairs for O/E/H). Each unique utterance embedding is extracted once and cached, with VoxCeleb1-H reusing the VoxCeleb1-E cache; ONNX models run through ONNX Runtime with the CUDA execution provider. The benchmark stores checksums, embeddings, per-trial scores, aggregate tables, and an environment snapshot, and is covered by unit tests for trial parsing, EER computation, and cache reuse\ifappendix\ (appendix Sec.~\ref{app:protocol}, \ref{app:repro})\fi.

\subsection{Verification results}
\label{ssec:verification}
Table~\ref{tab:main_results} reports the development path from the public ReDimNet2 checkpoint (1.601\% EER$_o$, 2.42\% EER$_p$ under our protocol). Direct fine-tuning on VoxBlink2 alone improves EER$_{ph}$ but is worse on clean VoxCeleb1, suggesting a difficult adaptation stage; the cosine-scheduled margin run also remains below the clean baseline. The first clear improvement appears when the training data is expanded and the input window is lengthened: adding VoxCeleb2, the backbone's pretraining corpus, approximately recovers the baseline pooled EER, and the multi-domain pool with 3D-Speaker, CN-Celeb, and CN-Celeb2 reduces EER$_p$ to 1.587\% and EER$_{ph}$ to 3.702\%. TidyVoice, KeSpeech, and longer training bring the best pre-LMFT checkpoint to 0.792\% EER$_o$ and 1.424\% EER$_p$. Applying LMFT directly to the public checkpoint gives 0.436\% EER$_o$ and 0.885\% EER$_p$. Starting LMFT from ReDimNet2+ pretrained further reduces EER$_o$ to 0.351\%, EER$_h$ to 1.055\%, EER$_p$ to 0.824\%, EER$_{ph}$ from the baseline's 7.214\% to 1.991\%, and raises Pr@$k$ from 0.5978 to 0.7413. A power-law fit of EER against training speakers and utterances is monotonic on VoxCeleb1-O but less so on E/H, so the domain and quality of added data matter beyond raw scale\ifappendix\ (appendix Fig.~\ref{fig:scaling})\fi. \ifappendix A separate runtime comparison uses fixed 6-second segments (appendix Table~\ref{tab:inference}); those measurements use a different window length from the 4-second verification evaluation.\fi

\subsection{Comparison with other models}
\label{ssec:comparison}
\begin{table*}[!t]
\caption{VoxCeleb1-O/E/H EER (\%) comparison with other models. ``Published'' values use their source protocols (VoxCeleb2-dev training); ``local, 4~s'' values use our shared 4-second window (Sec.~\ref{ssec:protocol}). Bold: best local value among the systems in this table.\label{tab:comparison}}
\centering
\small
\setlength{\tabcolsep}{6pt}
\begin{tabular}{llrrrr}
\toprule
System & Source & Params & EER$_o$ & EER$_e$ & EER$_h$ \\
\midrule
ECAPA-TDNN C=512 \cite{ecapa_tdnn} & published & 6.2M & 1.01 & 1.24 & 2.32 \\
ECAPA-TDNN C=1024 \cite{ecapa_tdnn} & published & 14.7M & 0.87 & 1.12 & 2.12 \\
CAM++ \cite{redimnet2} & published & 7.2M & 0.71 & 0.85 & 1.66 \\
ECAPA2 \cite{ecapa2} & published & 27.0M & 0.34 & 0.52 & 0.99 \\
WavLM \cite{redimnet2} & published & 324M & 0.52 & 0.63 & 1.34 \\
W2V-BERT 2.0 \cite{redimnet2} & published & 587M & 0.38 & 0.51 & 1.06 \\
ReDimNet2-B6 \cite{redimnet2} (full utterance) & published & 12.3M & 0.29 & 0.52 & 0.99 \\
\midrule
ReDimNet2-B6 \cite{redimnet2} (public ckpt.) & local, 4~s & 12.3M & 1.601 & 1.725 & 3.039 \\
WeSpeaker CAM++ \cite{wespeaker_toolkit,hf_wespeaker_campplus} & local, 4~s & 7.2M & 0.787 & 0.928 & 1.824 \\
WeSpeaker ResNet34-LM \cite{hf_wespeaker_resnet34_lm} & local, 4~s & 6.6M & 0.814 & 0.933 & 1.679 \\
WeSpeaker ECAPA512-LM \cite{hf_wespeaker_ecapa512_lm} & local, 4~s & 6.2M & 0.877 & 1.071 & 1.968 \\
\midrule
ReDimNet2+ LMFT (ours) & local, 4~s & 12.3M & \textbf{0.351} & \textbf{0.523} & \textbf{1.055} \\
\bottomrule
\end{tabular}
\end{table*}

Table~\ref{tab:comparison} distinguishes published results from our local 4-second comparison. The published ReDimNet2-B6 result is 0.29 / 0.52 / 0.99\% EER on O/E/H using full utterances \cite{redimnet2}; the same model family evaluated locally with 4-second windows gives 1.601 / 1.725 / 3.039\%. Restricting the input limits the available speaker evidence, so these protocols are not directly comparable. The window difference is a relevant source of the discrepancy, but its contribution is not isolated by these results. Under the shared 4-second protocol, ReDimNet2+ LMFT reaches 0.351 / 0.523 / 1.055\%, improving on the local starting checkpoint and all three evaluated WeSpeaker checkpoints. External systems differ in training corpora and budgets; this is a comparison of final systems under a shared evaluation protocol, not an equal-data architecture comparison.

\subsection{Retrieval reranking}
\label{ssec:rerank_results}
\begin{table}[!t]
\caption{Reranking ablation on ReDimNet2+ pretrained embeddings on two retrieval datasets. Pr@1/10/45 are measured on VoxCeleb1 (\%); VB2 is Pr@$k$ on the VoxBlink2 subset (fraction). Bold: best per column.\label{tab:rerank_results}}
\centering
\small
\setlength{\tabcolsep}{4pt}
\begin{tabular}{lrrrr}
\toprule
Method & Pr@1 & Pr@10 & Pr@45 & VB2 \\
\midrule
Baseline & \textbf{99.9577} & 99.8611 & 99.2404 & 0.7024 \\
Mean chain & \textbf{99.9577} & \textbf{99.9075} & 99.5299 & 0.7220 \\
Graph rerank & 99.9492 & 99.8798 & 99.5671 & 0.7369 \\
Mean chain + graph & 99.9511 & 99.9003 & \textbf{99.6992} & \textbf{0.7391} \\
\bottomrule
\end{tabular}
\end{table}

\begin{table}[!t]
\caption{VoxBlink2-subset retrieval Pr@$k$ before and after the full reranking pipeline (mean chain + graph, same settings) for three embedding models.\label{tab:rerank_transfer}}
\centering
\small
\begin{tabular}{lrrr}
\toprule
Embedding model & Before & After & $\Delta$ \\
\midrule
ReDimNet2 baseline & 0.5978 & 0.6906 & +0.0928 \\
ReDimNet2+ pretrained & 0.7024 & 0.7391 & +0.0367 \\
ReDimNet2+ LMFT & 0.7413 & \textbf{0.7687} & +0.0274 \\
\bottomrule
\end{tabular}
\end{table}

Table~\ref{tab:rerank_results} shows the reranking ablation on ReDimNet2+ pretrained embeddings across two datasets. On VoxCeleb1, Pr@1 is saturated at 99.95--99.96\% for every variant, so the remaining errors lie deeper in the list, where the two stages act: Pr@45 rises from 99.2404\% to 99.6992\%, i.e. the miss rate at $k{=}45$ falls from 0.76\% to 0.30\%, while on the VoxBlink2 subset Pr@$k$ rises from 0.7024 to 0.7391. Mean-chain reranking is conservative: it leaves Pr@1 unchanged, gives the best Pr@10 (99.9075\%), and raises VoxBlink2 Pr@$k$ to 0.7220. Graph reranking alone is more aggressive: it reaches a higher VoxBlink2 Pr@$k$ (0.7369) and Pr@45 (99.5671\%) but gives up a little Pr@1 and Pr@10, since the neighborhood terms can promote a candidate over the nearest one. The combination recovers most of that loss (99.9511\% and 99.9003\%) while giving the best Pr@45 and VoxBlink2 Pr@$k$, suggesting that the mean chain supplies cleaner neighbor lists to the graph stage.

Table~\ref{tab:rerank_transfer} applies the full pipeline with unchanged settings to three embedding models on the VoxBlink2 subset. The gain is largest for the weakest embedding (+0.093 for the public ReDimNet2 checkpoint) and smallest for the strongest (+0.027 after LMFT), as a better embedding leaves fewer ranking errors to repair. Reranking does not replace representation quality, however: the reranked baseline (0.6906) stays below the un-reranked ReDimNet2+ pretrained model (0.7024), and the LMFT gain (0.7024$\to$0.7413) and the reranking gain add up to the final 0.7687. Verification and retrieval thus need separate treatment: EER measures threshold behavior on labeled pairs, whereas Pr@$k$ measures ranking quality in a large gallery, and the graph stage improves the latter with structure invisible to pairwise scoring.

\section{Discussion and Conclusion}
\label{sec:conclusion}
ReDimNet2+ combines multi-corpus adaptation, waveform and codec augmentation, longer training windows, and LMFT. The development sequence improves verification under the shared 4-second evaluation protocol, while reranking improves retrieval on both VoxCeleb1 and a VoxBlink2 subset. Because several training factors change across runs, the development sequence does not isolate their individual effects. The predicted coloration shift motivates explicit codec and filtering augmentation; it does not establish that these transformations are necessary for every target domain. Failures remain under hard impostors and explicit degradation: EER$_h$ is about twice EER$_e$, and EER$_{ph}$ stays above the clean metrics after LMFT.

Limitations: external checkpoints are not controlled for training data, so the local comparison is shared-protocol rather than equal-data; EER$_{ph}$ uses synthetic degradations; and sub-1\% VoxCeleb1 EER alone does not prove in-the-wild robustness. Future work should target real telephone and VoIP data, calibration under degraded channels, and more out-of-domain benchmarks.\ifappendix\ Appendix~\ref{app:repro} describes the recorded configurations, checksums, embeddings, scores, and environment snapshots.\fi

\section{Compliance with Ethical Standards}
\label{sec:ethics}
This research study was conducted retrospectively using human subject speech data made available in open access by the VoxCeleb1, VoxCeleb2, VoxBlink2, 3D-Speaker, CN-Celeb, CN-Celeb2, TidyVoice (Common Voice), and KeSpeech corpora, and using publicly released pretrained speaker embedding checkpoints. No new recordings of human subjects were collected for this work.

No funding was received for conducting this study. The authors have no relevant financial or nonfinancial interests to disclose.

\section{Acknowledgments}
\label{sec:ack}
Claude Fable 5 (Anthropic) was used only for polishing the text of this manuscript. All ideas, methods, and experiments are the authors' own.

\bibliographystyle{IEEEbib}
\bibliography{references}

@misc{redimnet2,
      title={ReDimNet2: Scaling Speaker Verification via Time-Pooled Dimension Reshaping}, 
      author={Ivan Yakovlev and Anton Okhotnikov},
      year={2026},
      eprint={2603.11841},
      note={arXiv:2603.11841},
      archivePrefix={arXiv},
      primaryClass={eess.AS},
      url={https://arxiv.org/abs/2603.11841}, 
}

@inproceedings{w2v2bert,
author = {Arun Babu and others},
  title     = {A Massively Multilingual Self-Supervised Speech Model},
  booktitle = {Proceedings of the International Conference on Machine Learning (ICML)},
  year      = {2022},
  pages     = {1473--1483}
}

@INPROCEEDINGS{10096848,
  author={Sun, Yao and Zhang, Hanyi and Wang, Longbiao and Lee, Kong Aik and Liu, Meng and Dang, Jianwu},
  booktitle={ICASSP 2023 - 2023 IEEE International Conference on Acoustics, Speech and Signal Processing (ICASSP)}, 
  title={Noise-Disentanglement Metric Learning for Robust Speaker Verification}, 
  year={2023},
  volume={},
  number={},
  pages={1-5},
  doi={10.1109/ICASSP49357.2023.10096848}}

@inproceedings{ecapa_tdnn,
  title={{ECAPA-TDNN: Emphasized Channel Attention, Propagation and Aggregation in TDNN Based Speaker Verification}},
  author={Desplanques, Brecht and Thienpondt, Jenthe and Demuynck, Kris},
  booktitle={Interspeech 2020},
  pages={3830--3834},
  year={2020},
  doi={10.21437/Interspeech.2020-2650},
  url={https://arxiv.org/abs/2005.07143}
}

@misc{ecapa2,
      title={{ECAPA2: A Hybrid Neural Network Architecture and Training Strategy for Robust Speaker Embeddings}},
      author={Jenthe Thienpondt and Kris Demuynck},
      year={2024},
      eprint={2401.08342},
      note={arXiv:2401.08342},
      archivePrefix={arXiv},
      primaryClass={eess.AS},
      url={https://arxiv.org/abs/2401.08342}
}

@article{Mao2020ShortTimeSV,
  title        = {Short-time speaker verification with different speaking style utterances},
  author       = {Mao, Hongwei and Shi, Yan and Liu, Yue and Wei, Linqiang and Li, Yijie and Long, Yanhua},
  journal      = {PLOS ONE},
  year         = {2020},
  volume       = {15},
  number       = {11},
  pages        = {e0241809},
  doi          = {10.1371/journal.pone.0241809},
  pmcid        = {PMC7657545},
  pmid         = {33175898},
  url          = {https://doi.org/10.1371/journal.pone.0241809}
}

@misc{sphereface,
      title={Exploring Binary Classification Loss For Speaker Verification}, 
      author={Bing Han and Zhengyang Chen and Yanmin Qian},
      year={2023},
      eprint={2307.08205},
      note={arXiv:2307.08205},
      archivePrefix={arXiv},
      primaryClass={eess.AS},
      url={https://arxiv.org/abs/2307.08205}, 
}

@misc{li2026enhancingspeakerverificationw2vbert,
      title={Enhancing Speaker Verification with w2v-BERT 2.0 and Knowledge Distillation guided Structured Pruning}, 
      author={Ze Li and Ming Cheng and Ming Li},
      year={2026},
      eprint={2510.04213},
      note={arXiv:2510.04213},
      archivePrefix={arXiv},
      primaryClass={eess.AS},
      url={https://arxiv.org/abs/2510.04213}, 
}

@inproceedings{nisqa,
  title     = {{NISQA: A Deep CNN-Self-Attention Model for Multidimensional Speech Quality Prediction with Crowdsourced Datasets}},
  author    = {Gabriel Mittag and Babak Naderi and Assmaa Chehadi and Sebastian Möller},
  year      = {2021},
  booktitle = {{Interspeech 2021}},
  pages     = {2127--2131},
  doi       = {10.21437/Interspeech.2021-299},
  issn      = {2958-1796},
}

@INPROCEEDINGS{simamresnet,
  author={Qin, Xiaoyi and Li, Na and Weng, Chao and Su, Dan and Li, Ming},
  booktitle={ICASSP 2022 - 2022 IEEE International Conference on Acoustics, Speech and Signal Processing (ICASSP)}, 
  title={Simple Attention Module Based Speaker Verification with Iterative Noisy Label Detection}, 
  year={2022},
  volume={},
  number={},
  pages={6722-6726},
  doi={10.1109/ICASSP43922.2022.9746294}}

@inproceedings{vb2,
  title     = {{VoxBlink2: A 100K+ Speaker Recognition Corpus and the Open-Set Speaker-Identification Benchmark}},
  author    = {Yuke Lin and Ming Cheng and Fulin Zhang and Yingying Gao and Shilei Zhang and Ming Li},
  year      = {2024},
  booktitle = {{Interspeech 2024}},
  pages     = {4263--4267},
  doi       = {10.21437/Interspeech.2024-1490},
  issn      = {2958-1796},
}

@INPROCEEDINGS{cnceleb,
author = {Fan, Y. and others},
  booktitle={ICASSP 2020 - 2020 IEEE International Conference on Acoustics, Speech and Signal Processing (ICASSP)}, 
  title={CN-Celeb: A Challenging Chinese Speaker Recognition Dataset}, 
  year={2020},
  volume={},
  number={},
  pages={7604-7608},
  doi={10.1109/ICASSP40776.2020.9054017}}

@inproceedings{vc1,
  title     = {{VoxCeleb: A Large-Scale Speaker Identification Dataset}},
  author    = {Arsha Nagrani and Joon Son Chung and Andrew Zisserman},
  year      = {2017},
  booktitle = {{Interspeech 2017}},
  pages     = {2616--2620},
  doi       = {10.21437/Interspeech.2017-950},
  issn      = {2958-1796},
}

@misc{tidy,
      title={TidyVoice: A Curated Multilingual Dataset for Speaker Verification Derived from Common Voice}, 
      author={Aref Farhadipour and Jan Marquenie and Srikanth Madikeri and Eleanor Chodroff},
      year={2026},
      eprint={2601.16358},
      note={arXiv:2601.16358},
      archivePrefix={arXiv},
      primaryClass={eess.AS},
      url={https://arxiv.org/abs/2601.16358}, 
}

@inproceedings{kespeech,
  title     = {KeSpeech: An Open Source Speech Dataset of Mandarin and Its Eight Subdialects},
author = {Tang, Zhiyuan and others},
  booktitle = {NeurIPS 2021 Datasets and Benchmarks Track},
  year      = {2021},
  url       = {https://datasets-benchmarks-proceedings.neurips.cc/paper/2021/hash/0336dcbab05b9d5ad24f4333c7658a0e-Abstract-round2.html}
}

@article{cnceleb2,
author = {Li, Lantian and others},
    title = {CN-Celeb: Multi-genre speaker recognition},
    year = {2022},
    issue_date = {Feb 2022},
    volume = {137},
    number = {C},
    issn = {0167-6393},
    url = {https://doi.org/10.1016/j.specom.2022.01.002},
    doi = {10.1016/j.specom.2022.01.002},
    journal = {Speech Commun.},
    pages = {77--91},
    numpages = {15}
}

@misc{3d_speaker,
      title={3D-Speaker: A Large-Scale Multi-Device, Multi-Distance, and Multi-Dialect Corpus for Speech Representation Disentanglement}, 
      author={Siqi Zheng and Luyao Cheng and Yafeng Chen and Hui Wang and Qian Chen},
      year={2023},
      eprint={2306.15354},
      note={arXiv:2306.15354},
      archivePrefix={arXiv},
      primaryClass={cs.CL},
      url={https://arxiv.org/abs/2306.15354}, 
}

@InProceedings{Yun2019CutMix,
  author    = {Sangdoo Yun and Dongyoon Han and Seong Joon Oh and Sanghyuk Chun and Junsuk Choe and Youngjoon Yoo},
  title     = {CutMix: Regularization Strategy to Train Strong Classifiers with Localizable Features},
  booktitle = {The IEEE International Conference on Computer Vision ({ICCV})},
  year      = {2019},
  url       = {https://openaccess.thecvf.com/content\_ICCV\_2019/papers/Yun\_CutMix\_Regularization\_Strategy\_to\_Train\_Strong\_Classifiers\_With\_Localizable\_Features\_ICCV\_2019\_paper.pdf}
}

@inproceedings{park19e_interspeech,
  title     = {{SpecAugment: A Simple Data Augmentation Method for Automatic Speech Recognition}},
author = {Daniel S. Park and others},
  year      = {2019},
  booktitle = {{Interspeech 2019}},
  pages     = {2613--2617},
  doi       = {10.21437/Interspeech.2019-2680},
  issn      = {2958-1796},
}

@inproceedings{vc2,
  title     = {{VoxCeleb2: Deep Speaker Recognition}},
  author    = {Joon Son Chung and Arsha Nagrani and Andrew Zisserman},
  year      = {2018},
  booktitle = {{Interspeech 2018}},
  pages     = {1086--1090},
  doi       = {10.21437/Interspeech.2018-1929},
  issn      = {2958-1796},
}

@ARTICLE{wavlm,
author = {Chen, Sanyuan and others},
  journal={IEEE Journal of Selected Topics in Signal Processing}, 
  title={WavLM: Large-Scale Self-Supervised Pre-Training for Full Stack Speech Processing}, 
  year={2022},
  volume={16},
  number={6},
  pages={1505-1518},
  doi={10.1109/JSTSP.2022.3188113}}

@inproceedings{Zhong2017reranking,
  title     = {Re-ranking Person Re-identification with k-Reciprocal Encoding},
  author    = {Zhong, Zhun and Zheng, Liang and Cao, Donglin and Li, Shaozi},
  booktitle = {Proceedings of the IEEE Conference on Computer Vision and Pattern Recognition (CVPR)},
  pages     = {1318--1327},
  year      = {2017}
}

@inproceedings{suzuki2013centering,
  title        = {Centering Similarity Measures to Reduce Hubs},
  author       = {Suzuki, Ikumi and Hara, Kazuo and Shimbo, Masashi and Saerens, Marco and Fukumizu, Kenji},
  booktitle    = {Proceedings of the 2013 Conference on Empirical Methods in Natural Language Processing},
  pages        = {613--623},
  year         = {2013},
}

@inproceedings{w2v2,
  title     = {{Unsupervised Cross-Lingual Representation Learning for Speech Recognition}},
  author    = {Alexis Conneau and Alexei Baevski and Ronan Collobert and Abdelrahman Mohamed and Michael Auli},
  year      = {2021},
  booktitle = {{Interspeech 2021}},
  pages     = {2426--2430},
  doi       = {10.21437/Interspeech.2021-329},
  issn      = {2958-1796},
}

@INPROCEEDINGS{rir,
  author={Ko, Tom and Peddinti, Vijayaditya and Povey, Daniel and Seltzer, Michael L. and Khudanpur, Sanjeev},
  booktitle={2017 IEEE International Conference on Acoustics, Speech and Signal Processing (ICASSP)}, 
  title={A study on data augmentation of reverberant speech for robust speech recognition}, 
  year={2017},
  volume={},
  number={},
  pages={5220-5224},
  doi={10.1109/ICASSP.2017.7953152}}

@misc{musan,
  author = {David Snyder and Guoguo Chen and Daniel Povey},
  title = {{MUSAN}: {A} {M}usic, {S}peech, and {N}oise {C}orpus},
  year = {2015},
  eprint = {1510.08484},
  note = {arXiv:1510.08484v1}
}

@misc{li2026languageinvariantmultilingualspeakerverification,
      title={Language-Invariant Multilingual Speaker Verification for the TidyVoice 2026 Challenge}, 
      author={Ze Li and Xiaoxiao Miao and Juan Liu and Ming Li},
      year={2026},
      eprint={2603.08092},
      note={arXiv:2603.08092},
      archivePrefix={arXiv},
      primaryClass={eess.AS},
      url={https://arxiv.org/abs/2603.08092}, 
}

@misc{wespeaker_toolkit,
      title={Wespeaker: A Research and Production Oriented Speaker Embedding Learning Toolkit},
author = {Hongji Wang and others},
      year={2022},
      eprint={2210.17016},
      note={arXiv:2210.17016},
      archivePrefix={arXiv},
      primaryClass={eess.AS},
      url={https://arxiv.org/abs/2210.17016}
}

@misc{hf_wespeaker_resnet34_lm,
      title={{Wespeaker/wespeaker-voxceleb-resnet34-LM}},
      author={{Wespeaker}},
      year={2024},
      howpublished={Hugging Face model card},
      url={https://huggingface.co/Wespeaker/wespeaker-voxceleb-resnet34-LM}
}

@misc{hf_wespeaker_ecapa512_lm,
      title={{Wespeaker/wespeaker-ecapa-tdnn512-LM}},
      author={{Wespeaker}},
      year={2024},
      howpublished={Hugging Face model card},
      url={https://huggingface.co/Wespeaker/wespeaker-ecapa-tdnn512-LM}
}

@misc{hf_wespeaker_campplus,
      title={{Wespeaker/wespeaker-voxceleb-campplus}},
      author={{Wespeaker}},
      year={2024},
      howpublished={Hugging Face model card},
      url={https://huggingface.co/Wespeaker/wespeaker-voxceleb-campplus}
}

\ifappendix
\clearpage
\twocolumn[{\begin{center}
{\large\bf SUPPLEMENTARY MATERIAL (NOT PART OF THE ICASSP SUBMISSION)}\\[4pt]
Extended sections, tables, and figures from the journal-length version of this manuscript.
\end{center}\vspace{8pt}}]

\section{Related Work}
\label{app:related}
\subsection{Speaker Embedding Backbones}
Speaker verification systems have evolved from i-vector style pipelines to deep neural embeddings trained with metric or classification losses. Recent architectures combine convolutional front ends, temporal pooling, attention, and margin-based objectives to obtain compact speaker representations. ReDimNet2 follows this line by reshaping dimensions over time and pooling speaker evidence efficiently \citeapp{redimnet2}. SimAM-ResNet and related attention-based systems show that carefully designed convolutional backbones remain competitive for ASV \citeapp{simamresnet}. Larger self-supervised speech encoders such as WavLM and w2v-BERT provide general-purpose representations \citeapp{wavlm,w2v2bert,w2v2} at substantially higher inference cost, and often require distillation or pruning for efficient deployment \citeapp{li2026enhancingspeakerverificationw2vbert}. Our work keeps the ReDimNet2 backbone architecture fixed and asks to what extent robustness can be improved through data, augmentation, and fine-tuning alone.

\subsection{Robustness and Augmentation}
Channel and noise robustness have been studied through additive noise, reverberation, data augmentation, and metric learning \citeapp{10096848,Mao2020ShortTimeSV}. MUSAN and simulated room impulse responses are common ingredients for speaker and speech robustness recipes \citeapp{musan,rir}. SpecAugment and CutMix provide feature-level regularization \citeapp{park19e_interspeech,Yun2019CutMix}. In our setting, a shift in predicted coloration motivates codec simulation and filtering alongside standard noise and reverberation augmentation.

\subsection{Margin-Based Optimization}
Modern ASV systems commonly use angular or additive-margin objectives to increase inter-speaker separation. ReDimNet2 and SphereFace-style losses motivate the use of margins for speaker verification \citeapp{redimnet2,sphereface}. Recent multilingual speaker verification systems also use staged training and margin schedules to stabilize adaptation over heterogeneous data \citeapp{li2026languageinvariantmultilingualspeakerverification}. In our development sequence, direct fine-tuning with a high fixed margin degrades clean verification, while broad adaptation followed by LMFT improves it. Other training settings also change, so the sequence does not isolate margin scheduling.

\subsection{Retrieval Reranking}
Verification metrics evaluate binary trial decisions, but open-set identification depends on nearest-neighbor ranking. Reranking can improve retrieval by using local graph structure rather than independent pairwise scores. Reciprocal-neighbor reranking has been effective in person re-identification \citeapp{Zhong2017reranking}, and hubness-aware similarity correction addresses the tendency of some points to appear spuriously close to many queries \citeapp{suzuki2013centering}. We adapt these ideas to speaker embeddings through a lightweight graph reranking stage.

\section{Task and Metrics}
\label{app:metrics}
\subsection{Verification}
All models use randomly selected 4-second input windows for the reported verification evaluation. Checkpoints are selected on the VoxCeleb1 development set. Let $x_i$ be an input window and $f_\theta(x_i) \in \mathbb{R}^d$ be the speaker embedding extracted by a model. We L2-normalize embeddings and score a trial pair $(i,j)$ by cosine similarity:
\begin{equation}
s(i,j) = \frac{f_\theta(x_i)^T f_\theta(x_j)}
{\|f_\theta(x_i)\|_2\|f_\theta(x_j)\|_2}.
\end{equation}
Given labels $y(i,j)\in\{0,1\}$, the decision threshold controls false acceptance rate (FAR) and false rejection rate (FRR). Equal error rate (EER) is the operating point where FAR and FRR are equal or closest under the discrete score set. We report EER on VoxCeleb1-O, VoxCeleb1-E, and VoxCeleb1-H, as well as pooled EER, denoted EER$_p$, over the union of O/E/H scores.

\subsection{Robustness Stress Test}
Standard VoxCeleb1 protocols are useful but do not isolate robustness to channel and codec distortion. We therefore use an additional stress-test metric, EER$_{ph}$. Starting from the VoxCeleb1-O trial protocol, we generate 26 degraded variants of the same utterances using waveform and codec transformations. Scores from all degraded variants are concatenated and a single EER is computed over the resulting pooled list. By construction, the metric is severe: it measures whether a system maintains speaker separation when the same verification protocol is observed through many channel transformations simultaneously.

\subsection{Speaker Retrieval}
Speaker retrieval is evaluated separately on VoxCeleb1 and a VoxBlink2 subset using Precision@$k$. For a query embedding $e_i$, a gallery is ranked by cosine similarity and the metric is the fraction of relevant items among the first $k$ positions, averaged over queries. The Pr@1, Pr@10, and Pr@45 columns of Table~\ref{tab:rerank_results} are measured on VoxCeleb1 and expressed in percent. Its VB2 column and Table~\ref{tab:rerank_transfer} report VoxBlink2-subset Pr@$k$ as a fraction. Thus values near 99.9\% and 0.7 describe different evaluation datasets. This VoxBlink2-subset retrieval experiment is distinct from the published open-set identification benchmark \citeapp{vb2}. Verification and retrieval are related but not identical: a model may produce low EER on balanced trial pairs yet still suffer from hubness or local-ranking errors in a large gallery. For this reason, we evaluate reranking separately from verification.

\section{Data}
\label{app:data}
\subsection{Training Corpora}
The final training pool combines seven corpora. Table~\ref{tab:data} summarizes the scale. VoxBlink2 and VoxCeleb2 provide large in-the-wild YouTube speech; 3D-Speaker adds device, distance, and dialect variation; CN-Celeb and CN-Celeb2 add multi-genre web speech; TidyVoice contributes multilingual read speech derived from Common Voice; and KeSpeech contributes Mandarin and subdialect recordings. Checkpoints are selected on the VoxCeleb1 development set. VoxCeleb1 is used for verification and retrieval evaluation.

\begin{table*}[!t]
\caption{Training Corpora Used by ReDimNet2+\label{tab:data}}
\centering
\footnotesize
\begin{tabular}{lrrrrl}
\toprule
Corpus & Speakers & Languages & Utterances & Hours & Main domain \\
\midrule
VoxBlink2 \citeapp{vb2} & 11,053 & $\sim$18 & 673,277 & 1,457.77 & YouTube / in-the-wild \\
VoxCeleb2 \citeapp{vc2} & 5,994 & Multi & 1,092,009 & 2,442.00 & YouTube / interviews \\
3D-Speaker \citeapp{3d_speaker} & 10,000 & Mandarin dialects & 579,013 & 1,124.52 & Multi-device, multi-distance \\
CN-Celeb \citeapp{cnceleb} & 997 & Mandarin & 130,109 & 273.73 & Multi-genre web video \\
CN-Celeb2 \citeapp{cnceleb2} & 1,996 & Mandarin & 529,485 & 1,090.00 & Web-crawled speech \\
TidyVoice \citeapp{tidy} & 6,657 & 62 & 527,484 & 744.60 & Read multilingual speech \\
KeSpeech \citeapp{kespeech} & 27,237 & Mandarin + 8 subdialects & 976,064 & 1,542.00 & Read speech \\
\midrule
Total & 63,934 & $\sim$77 & 4,607,441 & 8,674.62 & Mixed \\
\bottomrule
\end{tabular}
\end{table*}

\subsection{VoxBlink2 Subset Overview}
The exploratory analysis was performed on a VoxBlink2 subset. The training split contains 673,277 FLAC files from 11,053 speakers (approximately 1,458 hours); the evaluation subset, selected for testing the model on our speaker-retrieval task (Pr@$k$), contains 134,697 FLAC files (approximately 345 hours). All files are 16~kHz mono, so the training and evaluation audio interface is fixed to that format throughout.

Speaker counts are imbalanced. In the training split, 69 speakers have exactly one recording, 411 have two to five recordings, and 4,870 have between 76 and 100 recordings. This imbalance affects both training and analysis: heavily represented speakers can dominate batches unless sampling is controlled, and lightly represented speakers provide little evidence for estimating robustness. We therefore treat balanced trial metrics and retrieval metrics as complementary diagnostics.

\subsection{Duration Mismatch}
Table~\ref{tab:duration} reports duration buckets. Training utterances are shorter on average: 11.0\% are shorter than 2 seconds and 22.7\% are shorter than 3 seconds. The evaluation subset contains more 5--20 second clips. We use short 2--3 second windows during early training because they cover most training examples and increase sample diversity through random crops from long files. During LMFT, we increase the input length to 96,000 samples, corresponding to 6 seconds at 16~kHz, and use repeat padding for short files so the model observes more speaker evidence per embedding.

\begin{table}[!t]
\caption{Duration Distribution in the VoxBlink2 Subset\label{tab:duration}}
\centering
\footnotesize
\begin{tabular}{lrr}
\toprule
Duration bucket & Train & Evaluation \\
\midrule
$<2$ s & 11.0\% & 2.5\% \\
2--3 s & 11.7\% & 7.2\% \\
3--5 s & 22.4\% & 21.2\% \\
5--10 s & 31.7\% & 39.4\% \\
10--20 s & 16.7\% & 21.8\% \\
20--30 s & 3.9\% & 5.0\% \\
30--60 s & 2.4\% & 2.8\% \\
\bottomrule
\end{tabular}
\end{table}

\subsection{Coloration and Byte-Rate Shift}
We use FLAC bytes per second as a heuristic indicator of signal compressibility. It is not a direct measurement of bandwidth or codec degradation: encoding settings, bit depth, silence, and acoustic content also affect file size. Table~\ref{tab:bytes_col} shows a large shift: the median byte rate drops from 21,040 bytes/s in train to 14,021 bytes/s in the evaluation subset, and the fifth percentile drops from 16,283 to 8,989 bytes/s.

We also computed NISQA quality dimensions \citeapp{nisqa}. The coloration (COL) predictions provide complementary evidence of a quality shift associated with spectral distortion, bandwidth limitation, and codec artifacts. Only 6.2\% of train files have COL below 2.0, compared with 41.4\% of evaluation files. This observation motivates including filtering and codec simulation alongside noise and reverberation in the augmentation recipe of Appendix~\ref{app:pipeline}. It does not establish a unique physical cause for the shift or the necessity of a particular augmentation.

\begin{table}[!t]
\caption{Byte-Rate and Coloration Shift Between Train and the Pr@$k$ Evaluation Subset\label{tab:bytes_col}}
\centering
\footnotesize
\begin{tabular}{lrr}
\toprule
Statistic & Train & Evaluation \\
\midrule
Median bytes/s & 21,040 & 14,021 \\
p05 bytes/s & 16,283 & 8,989 \\
p95 bytes/s & 26,860 & 20,831 \\
Files with COL $<2.0$ & 41,599 (6.2\%) & 55,820 (41.4\%) \\
Median COL gap & \multicolumn{2}{c}{$-0.977$} \\
\bottomrule
\end{tabular}
\end{table}

\subsection{Within-Speaker Channel Variation}
For 884 speakers with at least five recordings, the median within-speaker standard deviation of bytes/s is 1,753. Only 16 speakers, or 1.8\%, have within-speaker standard deviation below 500 bytes/s, while 364 speakers, or 41\%, exceed 2,000 bytes/s. The variability is consistent with heterogeneous recording conditions within a speaker, although byte rate alone cannot establish channel variation or positive-pair difficulty. This motivates training representations that remain compact across recording conditions.

\section{Training Pipeline}
\label{app:pipeline}
\subsection{Manifest and Random-Window Decoding}
The pipeline begins with a Parquet manifest that stores audio metadata. At initialization, the loader builds an index over rows and a speaker-to-label mapping. During training, a sample is selected, a fixed-length time window is chosen, and only that range is decoded; on a decode failure, the loader falls back to a neighboring valid example rather than aborting the run. This behavior matters for large crawled corpora, where isolated problematic files should not interrupt multi-day training.

All audio is converted to mono 16~kHz. Short segments are repeat-padded to the requested number of samples; repeat padding preserves speech content and avoids long artificial silence tails during LMFT, which would otherwise distort the embedding distribution at long input lengths.

\subsection{Pipeline Speed}
Table~\ref{tab:pipeline_speed} summarizes key preprocessing benchmarks. Range decoding is 2.6$\times$ faster than decoding the full file and cropping afterward. Speed perturbation was optimized by restricting it to a fixed set of ten resampling factors in $[0.9,1.1]$, which allows kernel reuse and reduces per-sample overhead. Codec augmentation calls FFmpeg directly rather than through a higher-level Python wrapper. Finally, spectrogram computation was moved from the CPU collator to the GPU training step, which improved throughput when using multiple data-loader workers.

\begin{table}[!t]
\caption{Preprocessing and Feature-Extraction Speedups\label{tab:pipeline_speed}}
\centering
\footnotesize
\begin{tabular}{lrrr}
\toprule
Component & Baseline & Optimized & Speedup \\
\midrule
Random audio window & 5.38 ms & 2.06 ms & 2.6$\times$ \\
Speed perturbation & 455 ms & 0.16 ms & $>2800\times$ \\
PCM A-law codec & 14.33 ms & 8.77 ms & 1.63$\times$ \\
G.722 codec & 14.74 ms & 9.52 ms & 1.55$\times$ \\
MP3 8 kHz codec & 16.39 ms & 10.69 ms & 1.53$\times$ \\
Spectrograms, 4 workers & 1029 samp/s & 1570 samp/s & 1.53$\times$ \\
\bottomrule
\end{tabular}
\end{table}


\subsection{Waveform and Codec Augmentation}
Each training example is assigned to one of four top-level modes: keep, codec only, waveform augmentation only, or codec plus waveform augmentation. The probabilities used in the main recipe are 0.5, 0.2, 0.2, and 0.1, respectively. This mix preserves enough clean speech for stable classification while forcing the model to see codec and channel shifts frequently.

Waveform augmentation is organized by levels. Filters include band-pass, band-stop, high-pass, and low-pass transformations. Additive noise includes colored noise, background samples from MUSAN \citeapp{musan}, and RawBoost-style corruption. RIR augmentation uses simulated room impulse responses \citeapp{rir}. Speed perturbation samples a factor between 0.9 and 1.1. Table~\ref{tab:aug_policy} gives the high-level policy.

\begin{table}[!t]
\caption{Augmentation Policy Used During Training\label{tab:aug_policy}}
\centering
\scriptsize
\setlength{\tabcolsep}{3pt}
\begin{tabular}{lcp{0.42\columnwidth}}
\toprule
Stage & Probability & Transform family \\
\midrule
Top-level keep & 0.5 & No waveform corruption \\
Top-level codec & 0.2 & Codec branch only \\
Top-level augment & 0.2 & Waveform branch only \\
Top-level codec+augment & 0.1 & Both branches \\
\midrule
Filters & 0.3 & Band-pass, band-stop, high-pass, low-pass \\
Additive noise & 0.3 & Colored noise, MUSAN, RawBoost \\
RIR & 0.3 & Room impulse response \\
Perturb & 0.1 & Speed perturbation, 0.9--1.1 \\
Feature aug. & 0.2 & Frequency mask, time mask, CutMix \\
\bottomrule
\end{tabular}
\end{table}

Codec augmentation simulates compression and telephony-like channels. The preset set includes MP3, Opus, AAC, FLAC, Vorbis, G.723.1, IMA ADPCM WAV, G.722, A-law, $\mu$-law, Speex, and AMR-NB. Several presets are applied at both 8 and 16~kHz. This branch is directly motivated by the coloration shift in Appendix~\ref{app:data}. Feature-level augmentation is applied after GPU spectrogram computation and includes SpecAugment and CutMix \citeapp{park19e_interspeech,Yun2019CutMix}.

\begin{figure*}[!t]
\centering
\subfloat[]{\includegraphics[width=0.19\textwidth]{waveforms/00001__mel_clean.png}}
\hfil
\subfloat[]{\includegraphics[width=0.19\textwidth]{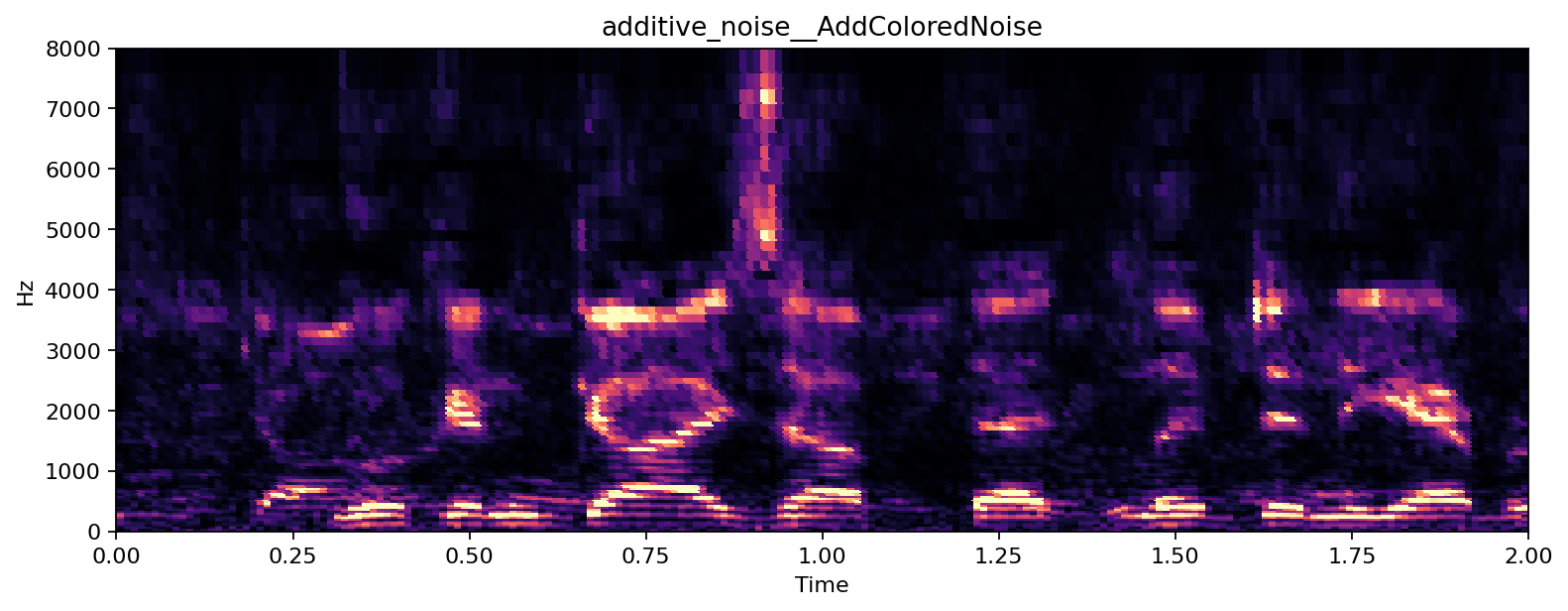}}
\hfil
\subfloat[]{\includegraphics[width=0.19\textwidth]{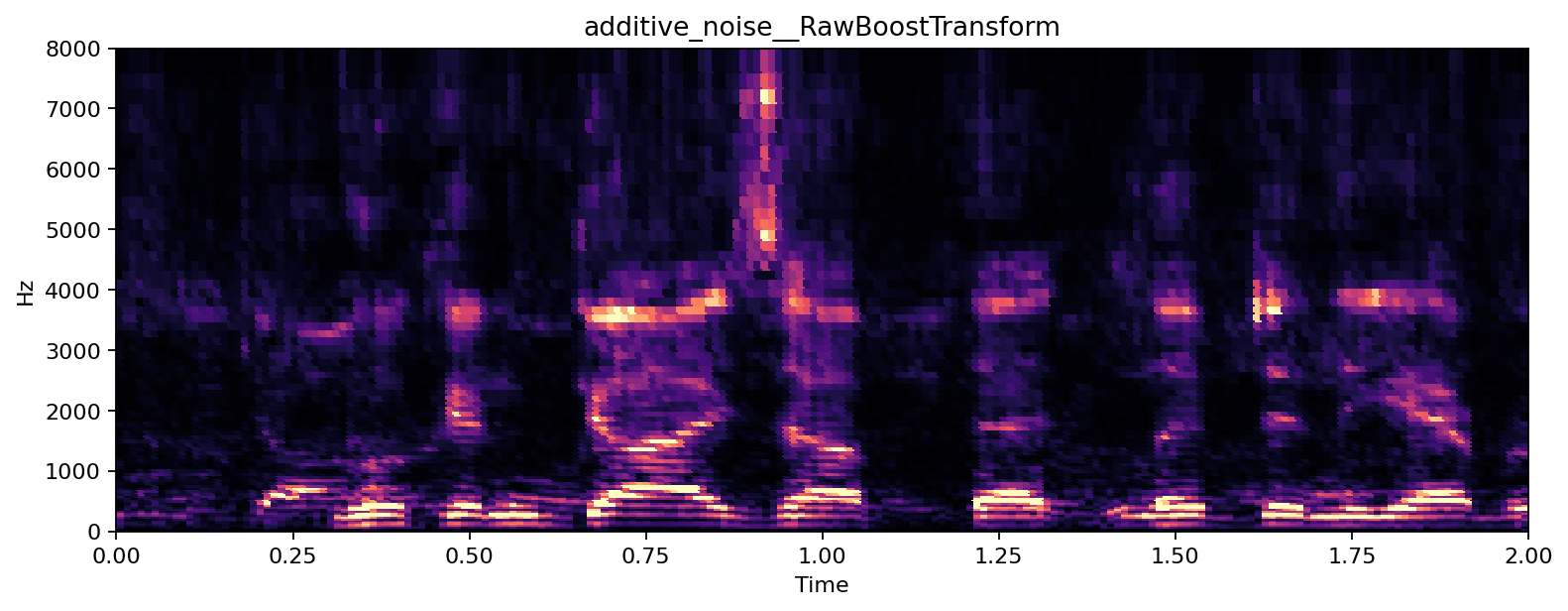}}
\hfil
\subfloat[]{\includegraphics[width=0.19\textwidth]{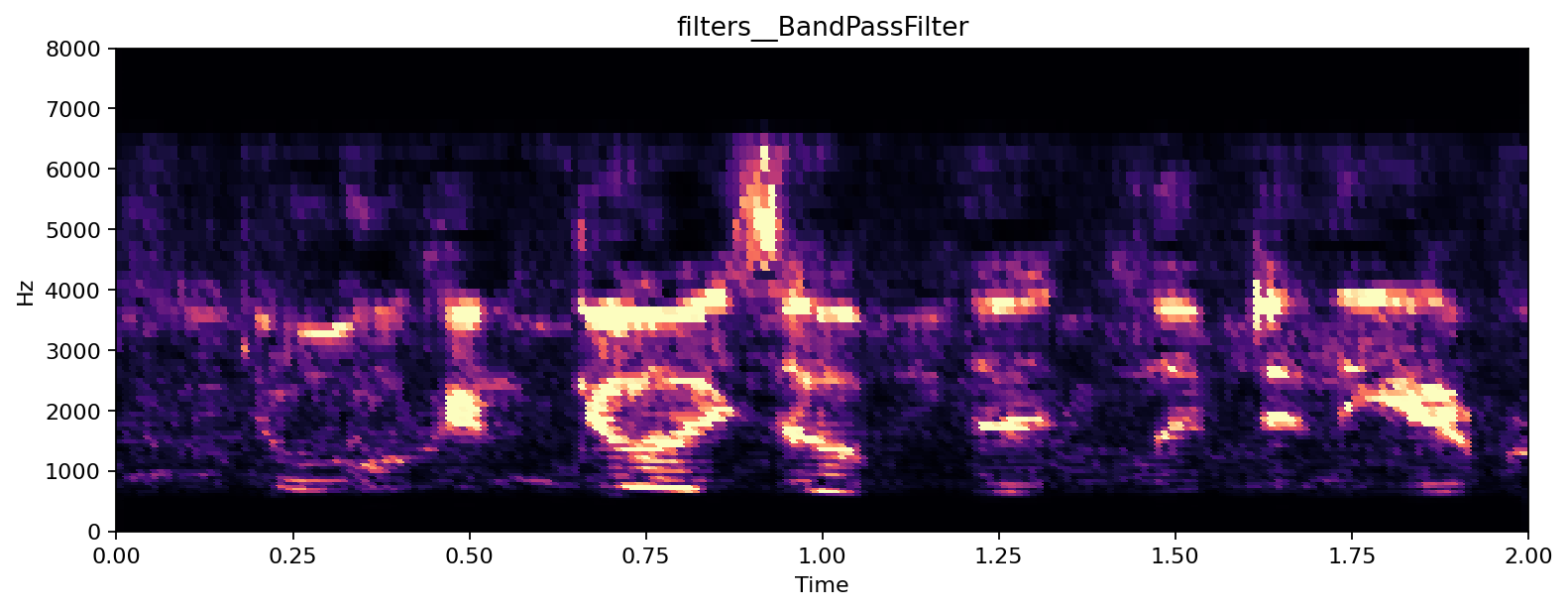}}
\hfil
\subfloat[]{\includegraphics[width=0.19\textwidth]{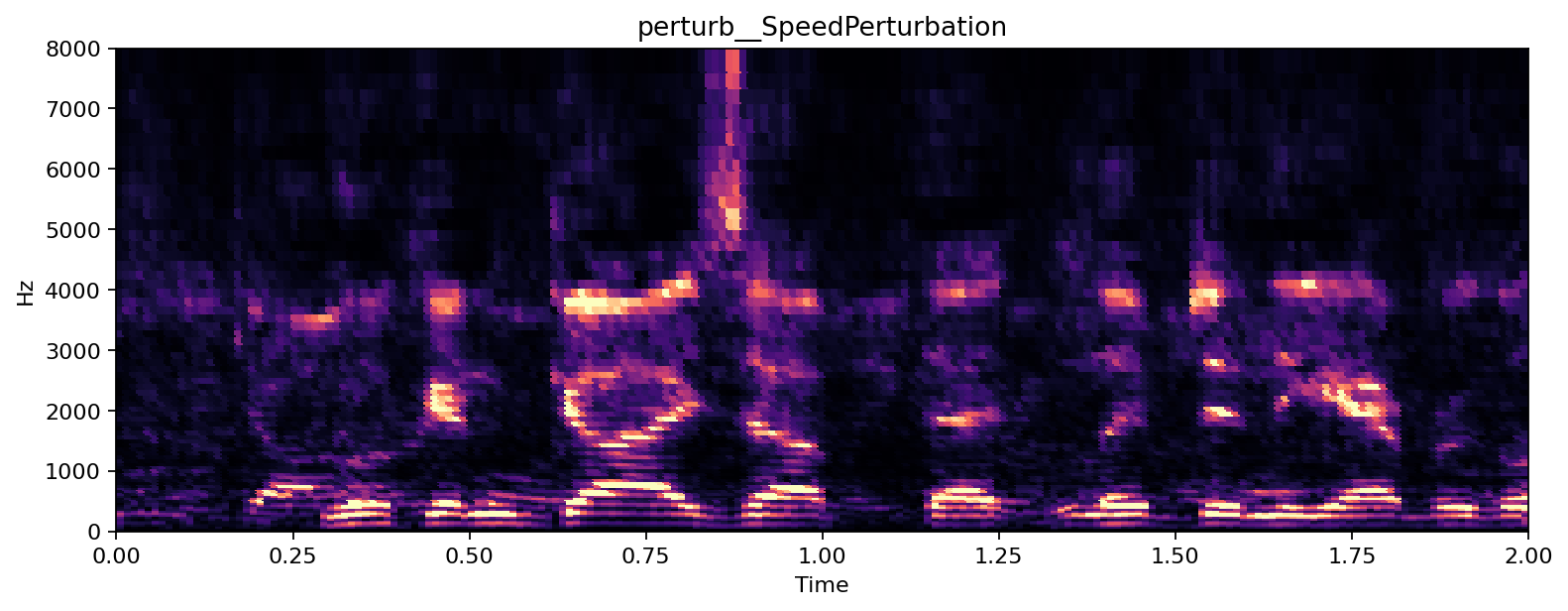}}
\caption{Examples of waveform-level transformations: clean mel reference, colored noise, RawBoost, band-pass filtering, and speed perturbation.}
\label{fig:waveform_aug}
\end{figure*}


\section{Model and Optimization}
\label{app:model}
\subsection{Backbone}
ReDimNet2 is a speaker representation network that alternates local aggregation and dimensional reshaping, allowing the model to capture time-frequency structure with moderate compute \citeapp{redimnet2}. We use the pretrained ReDimNet2 checkpoint as the starting point. The model operates over spectrogram features and outputs a fixed-dimensional embedding for each utterance. During scoring, embeddings are L2-normalized and compared by cosine similarity.

\subsection{Fine-Tuning Stages}
The baseline is the unmodified public ReDimNet2 checkpoint. Every adaptation run in Table~\ref{tab:main_results} starts independently from that checkpoint, except the final ReDimNet2+ LMFT run. Thus adjacent rows are not successive checkpoints from one continuous run. The rows with added corpora describe independent fine-tuning experiments with expanded training pools.

ReDimNet2+ pretrained is obtained by fine-tuning the public checkpoint on all seven training corpora. The \emph{ReDimNet2 (LMFT only)} control applies LMFT directly to the public checkpoint. In contrast, ReDimNet2+ LMFT starts from ReDimNet2+ pretrained and applies LMFT after multi-corpus adaptation. Checkpoints are selected on the VoxCeleb1 development set.

The direct fine-tuning run uses a learning rate of $5\times10^{-5}$, whereas the cosine-margin run uses $10^{-4}$ and a margin schedule from 0.0 to 0.2. Since training duration and learning rate also differ, their results do not isolate margin scheduling. Final LMFT uses a fixed margin of 0.3 and 6-second training windows.

\subsection{Speaker Classification Objective}
We use a SphereFace2 classification head \citeapp{sphereface}, which trains one-versus-all binary classifiers for normalized embeddings and normalized speaker-class weights. The objective follows the cited binary-classification formulation, rather than multiclass softmax normalization. Early experiments used either a fixed margin or a cosine schedule from 0.0 to 0.2. The final LMFT stage uses a fixed margin of 0.3 and 6-second training segments. All models use randomly selected 4-second windows for evaluation.

\subsection{Hyperparameters}
Table~\ref{tab:hparams} summarizes the main training runs. Training used distributed execution with Accelerate and FSDP, BFloat16 precision, AdamW, gradient clipping at 20, fixed seed 42, and checkpointed recovery. Batch size is reported per GPU; accumulation denotes gradient accumulation steps. Early runs used 32,200-sample windows. Later runs increased to 48,300 samples, and LMFT used 96,000 samples.

\begin{table*}[!t]
\caption{Training Hyperparameters for the Main Runs\label{tab:hparams}}
\centering
\footnotesize
\begin{tabular}{lrrrrrr}
\toprule
Run & Batch/GPU & GPUs & Accum. & Epochs & LR & Samples \\
\midrule
ReDimNet2 (FT) & 64 & 6 & 4 & 22 & $5\cdot10^{-5}$ & 32,200 \\
ReDimNet2 (cosine margin schedule) & 64 & 6 & 4 & 10 & $10^{-4}$ & 32,200 \\
ReDimNet2 (VC2, 6 epochs) & 64 & 6 & 4 & 6 & $9\cdot10^{-5}$ & 32,200 \\
ReDimNet2 (multi-domain, 6 epochs) & 48 & 6 & 5 & 6 & $9\cdot10^{-5}$ & 48,300 \\
ReDimNet2 (multi-domain, 10 epochs) & 48 & 6 & 5 & 10 & $9\cdot10^{-5}$ & 48,300 \\
ReDimNet2 (TidyVoice added) & 48 & 6 & 5 & 6 & $9\cdot10^{-5}$ & 48,300 \\
ReDimNet2 (KeSpeech added) & 48 & 6 & 5 & 6 & $9\cdot10^{-5}$ & 48,300 \\
ReDimNet2+ pretrained & 48 & 6 & 5 & 10 & $9\cdot10^{-5}$ & 48,300 \\
ReDimNet2+ LMFT & 24 & 6 & 10 & 6 & $10^{-4}$--$3\cdot10^{-5}$ & 96,000 \\
\bottomrule
\end{tabular}
\end{table*}

\section{External Benchmark Protocol}
\label{app:protocol}
\subsection{Model Selection}
We selected external baselines that are public, widely usable, and easy to run locally. WeSpeaker provides a research and production speaker embedding toolkit \citeapp{wespeaker_toolkit} and public ONNX checkpoints. We evaluated WeSpeaker ResNet34-LM \citeapp{hf_wespeaker_resnet34_lm}, WeSpeaker ECAPA512-LM \citeapp{hf_wespeaker_ecapa512_lm}, and WeSpeaker CAM++ \citeapp{hf_wespeaker_campplus}. These models cover different open-source speaker embedding families while keeping inference practical on the local GPU.

\subsection{Local VoxCeleb1 Setup}
All reported local systems were evaluated with randomly selected 4-second input windows on the same local VoxCeleb1 trial files. Checkpoints were selected on the VoxCeleb1 development set. The published ReDimNet2-B6 reference instead uses full utterances and cosine scoring without score normalization \citeapp{redimnet2}. A fixed window limits speaker evidence and makes our local numbers protocol-specific; the available results do not isolate how much of the published-to-local gap is due to duration. VoxCeleb1-O contains 37,611 trial pairs and references 4,708 unique utterances. VoxCeleb1-E contains 579,818 trial pairs and references 145,160 unique utterances. VoxCeleb1-H contains 550,894 trial pairs and references 137,924 unique utterances. The utterance set of VoxCeleb1-H is a subset of VoxCeleb1-E, so the benchmark extracts embeddings once for VoxCeleb1-E and reuses that embedding cache when scoring VoxCeleb1-H. Missing VoxCeleb1 test-speaker audio was filled from the Hugging Face VoxCeleb mirror. The downloaded test zip checksum was verified and all trial references resolved before scoring.

The benchmark script extracts each unique utterance embedding once, stores compressed embeddings, computes all trial cosine scores, writes a score CSV, and writes a JSON metric file. For WeSpeaker ONNX models, ONNX Runtime was configured with CUDA execution provider and CPU fallback. The experiment directory stores model files, checksums, logs, pip freeze, GPU snapshot, environment metadata, embeddings, scores, and aggregate tables.

\section{Reranking}
\label{app:reranking}
\subsection{Mean-Chain Reranking}
Let $k$ denote the evaluation cutoff, $K$ the graph neighborhood size, and $M>K$ the initial candidate-pool size. The first reranking stage works inside this candidate pool. For query embedding $e_i$, all embeddings are L2-normalized and an initial pool $P(i)$ is selected by cosine similarity. The probe vector starts as
\begin{equation}
p_i^{(0)} = e_i.
\end{equation}
At step $t$, the algorithm selects the unused candidate with highest similarity to the current probe:
\begin{equation}
j_t = \arg\max_{j\in P(i)\setminus\{j_1,\ldots,j_{t-1}\}} e_j^Tp_i^{(t-1)}.
\end{equation}
The probe is then updated as the normalized mean of the original query and the last selected neighbor:
\begin{equation}
p_i^{(t)} = \frac{e_i + e_{j_t}}{\|e_i+e_{j_t}\|_2}.
\end{equation}
After $K$ selections, the chain order defines the neighbor list $N(i)$. Each probe averages the original query with only the most recently selected neighbor; it is not a cumulative mean. Anchoring the probe to the query limits drift while allowing the order to follow local embedding structure.

\subsection{Graph Reranking}
The second stage builds a $K$-nearest-neighbor graph and uses its structure without speaker labels. For query $i$, the local candidate set contains current neighbors and neighbors of those neighbors:
\begin{equation}
P_{\mathrm{graph}}(i)=N(i)\cup \bigcup_{u\in N(i)}N(u), \quad i\notin P_{\mathrm{graph}}(i).
\end{equation}
Each candidate is scored by rank support, reciprocal support, common-neighbor support, and a hubness penalty:
\begin{equation}
S(i,j)=w_rR(i,j)+w_qQ(i,j)+w_cC(i,j)-w_hH(j),
\end{equation}
where the final recipe uses $w_r=1.0$, $w_q=1.5$, $w_c=2.0$, and $w_h=0.3$. The rank term rewards candidates already high in $N(i)$. The reciprocal term rewards candidates whose own neighbor list contains $i$. The common-neighbor term rewards candidates sharing local context with $i$. The hubness term penalizes embeddings with excessive in-degree:
\begin{equation}
H(j)=\max\left(0,\log\frac{\mathrm{deg}_{in}(j)}{K}\right).
\end{equation}
Here $\mathrm{deg}_{in}(j)$ counts the lists that contain candidate $j$. With $K$ outgoing neighbors per graph node, the mean in-degree is $K$; the penalty is zero up to that value and increases logarithmically above it. After rescoring, each list is replaced by its top-$K$ candidates and the graph is rebuilt. The final experiments use three iterations and unchanged weights across embedding models. Evaluation reads the requested top-$k$ positions from the resulting ranking.

\section{Results}
\label{app:results}
\subsection{Main Verification Results}
Table~\ref{tab:main_results} reports the main ReDimNet2+ development path under a shared 4-second evaluation protocol. The baseline ReDimNet2 checkpoint obtains 1.601\% EER on VoxCeleb1-O and 2.42\% pooled EER. The first direct fine-tuning run is worse on clean VoxCeleb1, which supports the hypothesis that a high fixed margin is too aggressive at the beginning of adaptation. Cosine-scheduled margin training is still not sufficient. The first clear improvement appears when the training data is expanded and the model receives more input samples. The best pre-LMFT checkpoint reaches 0.792\% EER$_o$ and 1.424\% EER$_p$. The LMFT-only control starts from the public checkpoint and reaches 0.436\% EER$_o$ and 0.885\% EER$_p$. Applying LMFT to ReDimNet2+ pretrained further reduces EER$_o$ to 0.351\%, EER$_h$ to 1.055\%, EER$_p$ to 0.824\%, and EER$_{ph}$ to 1.991\%.


\subsection{Published VoxCeleb1 Baselines}
Table~\ref{tab:published_baselines} situates the result among published VoxCeleb1 numbers. Systems differ in training corpora, model scale, calibration, and implementation details, so the comparison is indicative rather than equal-budget. ReDimNet2+ is competitive with strong VoxCeleb2-dev systems and trails the extended-data W2V-BERT 2.0 result reported in \citeapp{redimnet2}. The controlled contribution of this paper is the reproducible local comparison in Table~\ref{tab:external_results} rather than a universal state-of-the-art claim.

\begin{table*}[!t]
\caption{Published VoxCeleb1 results under their source protocols. ReDimNet2-B6 uses full utterances; our results use 4-second windows. The local checkpoint comparison is given in Table~\ref{tab:comparison}.\label{tab:published_baselines}}
\centering
\footnotesize
\begin{tabular}{llrrrr}
\toprule
System & Training data & Params & EER$_o$ & EER$_e$ & EER$_h$ \\
\midrule
ECAPA-TDNN C=512 \citeapp{ecapa_tdnn} & VoxCeleb2-dev & 6.2M & 1.01 & 1.24 & 2.32 \\
ECAPA-TDNN C=1024 \citeapp{ecapa_tdnn} & VoxCeleb2-dev & 14.7M & 0.87 & 1.12 & 2.12 \\
CAM++ \citeapp{redimnet2} & VoxCeleb2-dev & 7.2M & 0.71 & 0.85 & 1.66 \\
ECAPA2 \citeapp{ecapa2} & VoxCeleb2-dev & 27.0M & 0.34 & 0.52 & 0.99 \\
WavLM \citeapp{redimnet2} & VoxCeleb2-dev & 324M & 0.52 & 0.63 & 1.34 \\
W2V-BERT 2.0 \citeapp{redimnet2} & VoxCeleb2-dev & 587M & 0.38 & 0.51 & 1.06 \\
ReDimNet2-B6 \citeapp{redimnet2} (full utterance) & VoxCeleb2-dev & 12.3M & 0.29 & 0.52 & 0.99 \\
SimAM-ResNet100 \citeapp{redimnet2,simamresnet} & VoxBlink2+VoxCeleb2 & 50.2M & 0.23 & 0.46 & 0.87 \\
W2V-BERT 2.0 \citeapp{redimnet2} & VoxBlink2+VoxCeleb2 & 587M & \textbf{0.14} & \textbf{0.31} & \textbf{0.73} \\
\midrule
ReDimNet2+ LMFT (ours, 4~s) & Mixed corpora & 12.3M & 0.351 & 0.523 & 1.055 \\
\bottomrule
\end{tabular}
\end{table*}

\subsection{Data Scaling}
Fig.~\ref{fig:scaling} shows EER trends as the number of training speakers and utterances increases. We fit a power law in log--log space. VoxCeleb1-O follows the most stable monotonic trend, while VoxCeleb1-E and VoxCeleb1-H become less monotonic at the largest settings. This indicates that raw scale helps but is not the only variable: the domain and quality of added data influence the extended and hard protocols.

\begin{figure*}[!t]
\centering
\subfloat[]{\includegraphics[width=0.32\textwidth]{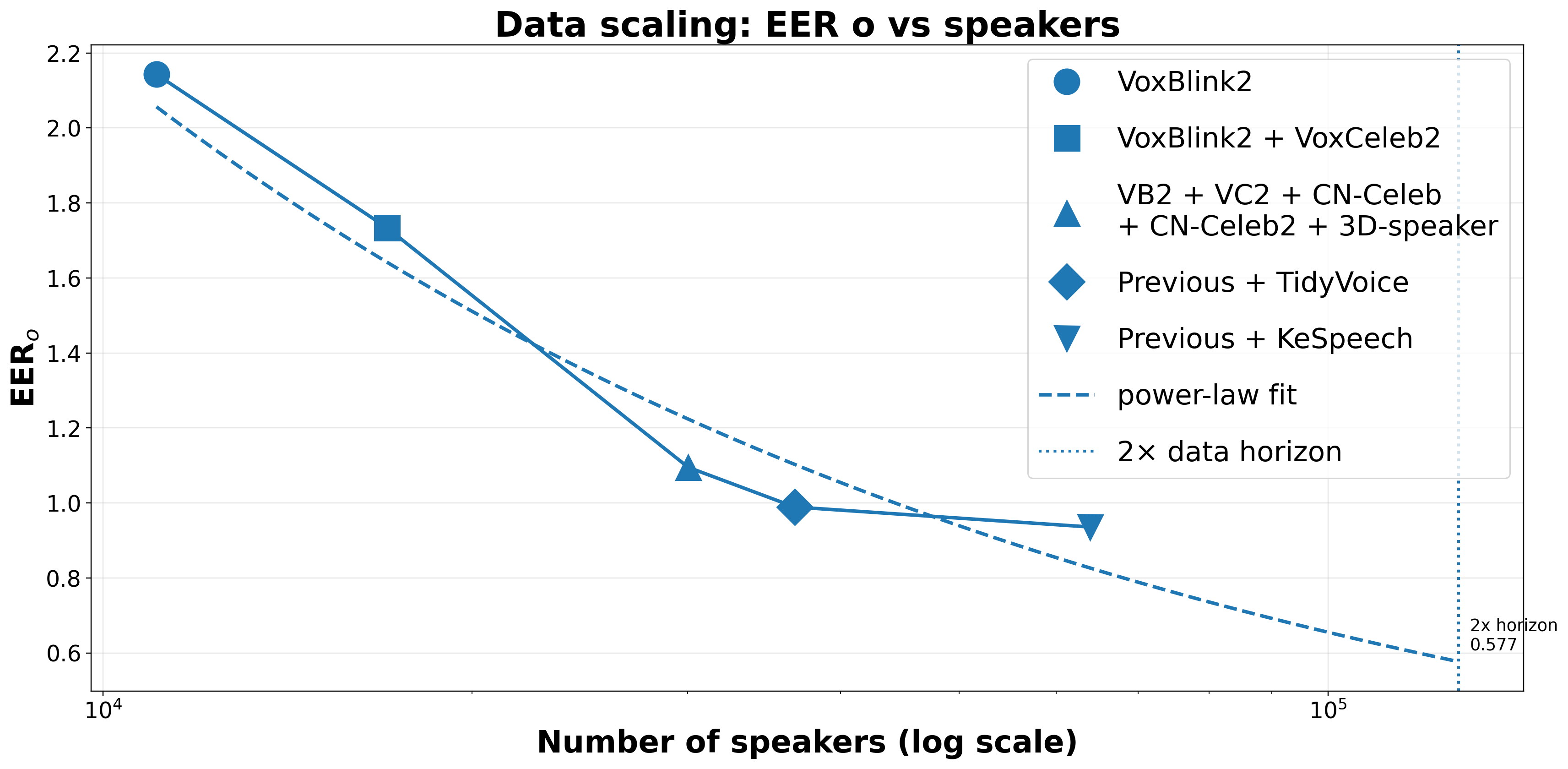}}
\hfil
\subfloat[]{\includegraphics[width=0.32\textwidth]{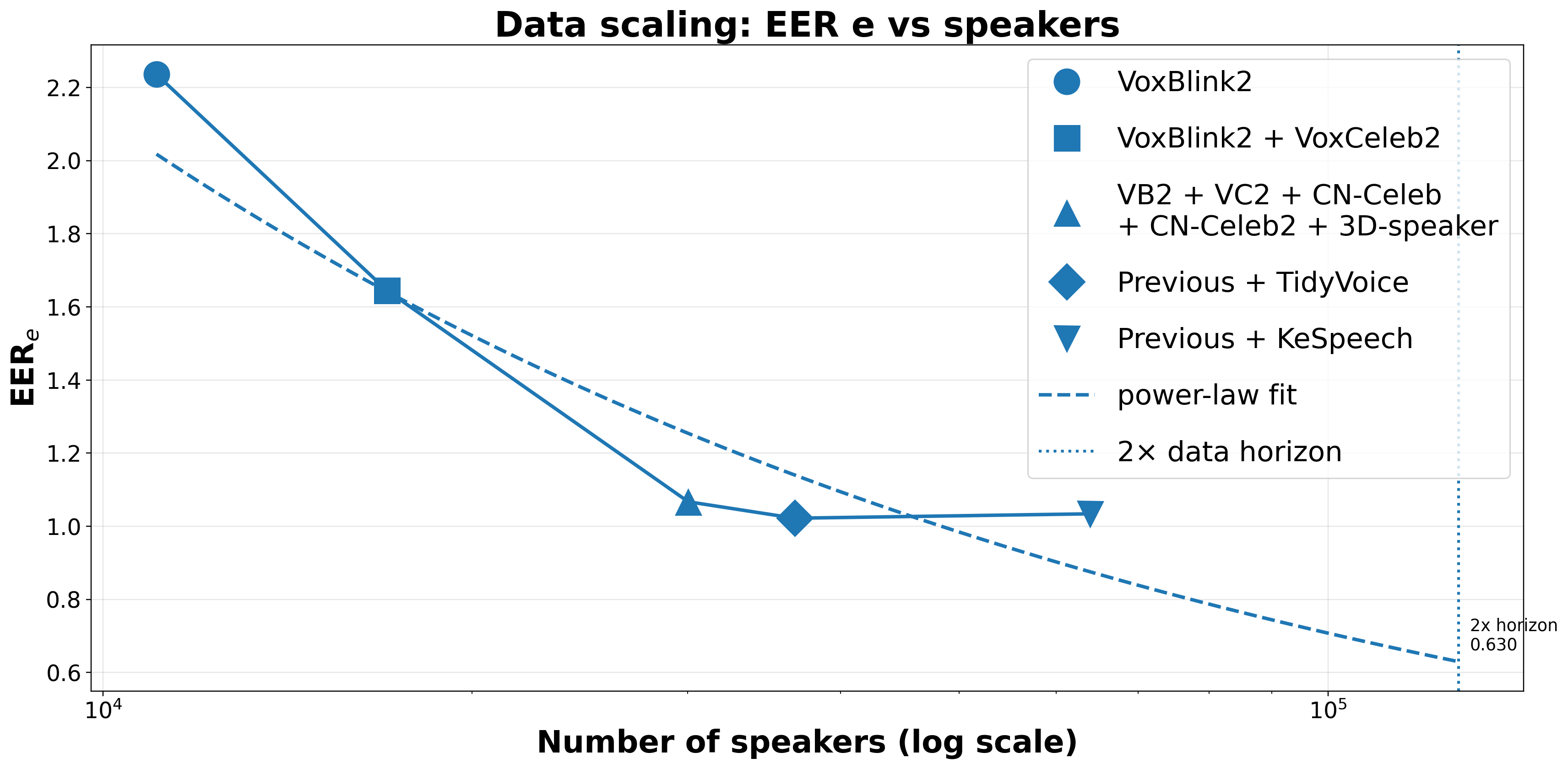}}
\hfil
\subfloat[]{\includegraphics[width=0.32\textwidth]{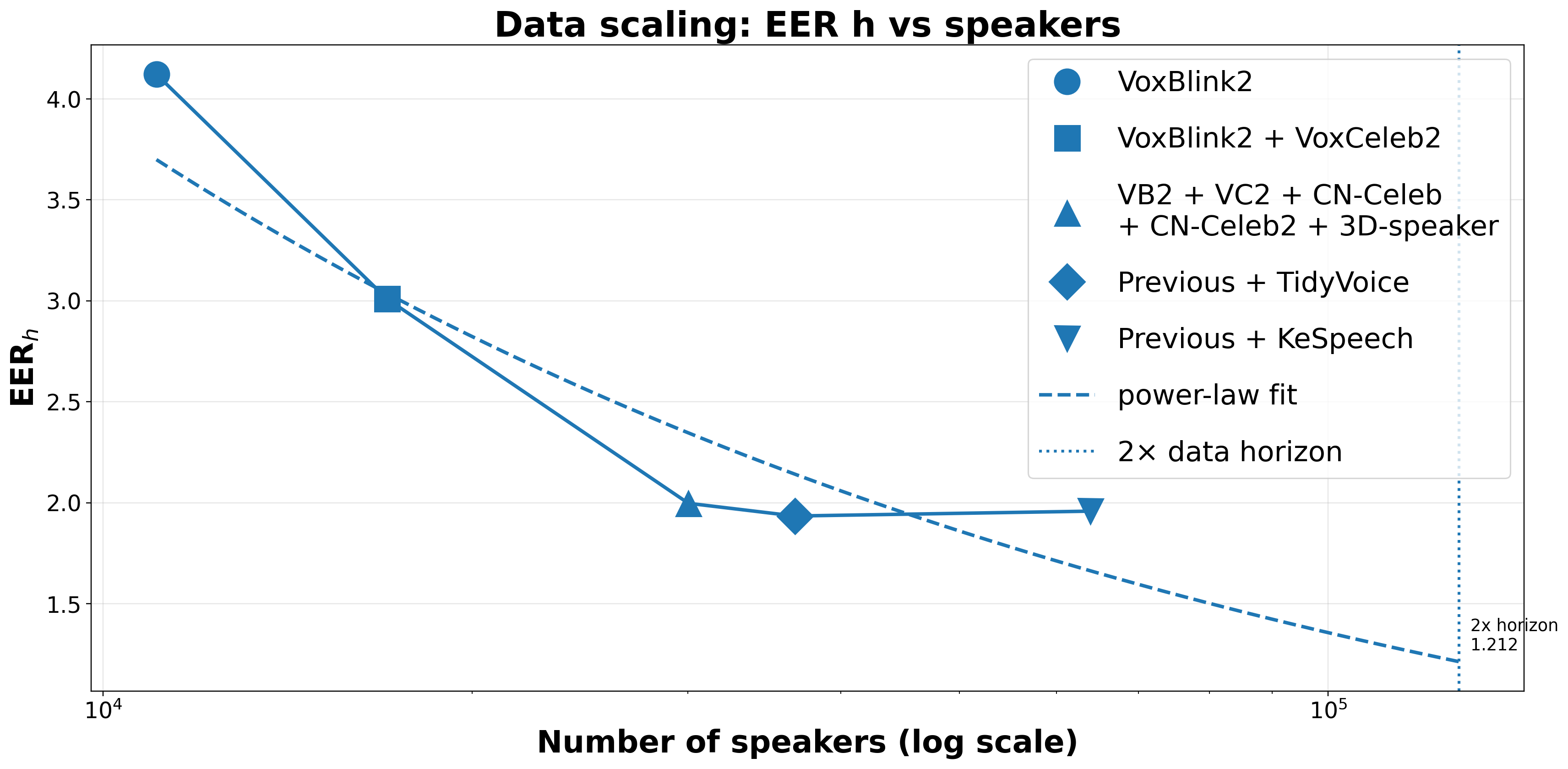}}
\hfil
\subfloat[]{\includegraphics[width=0.32\textwidth]{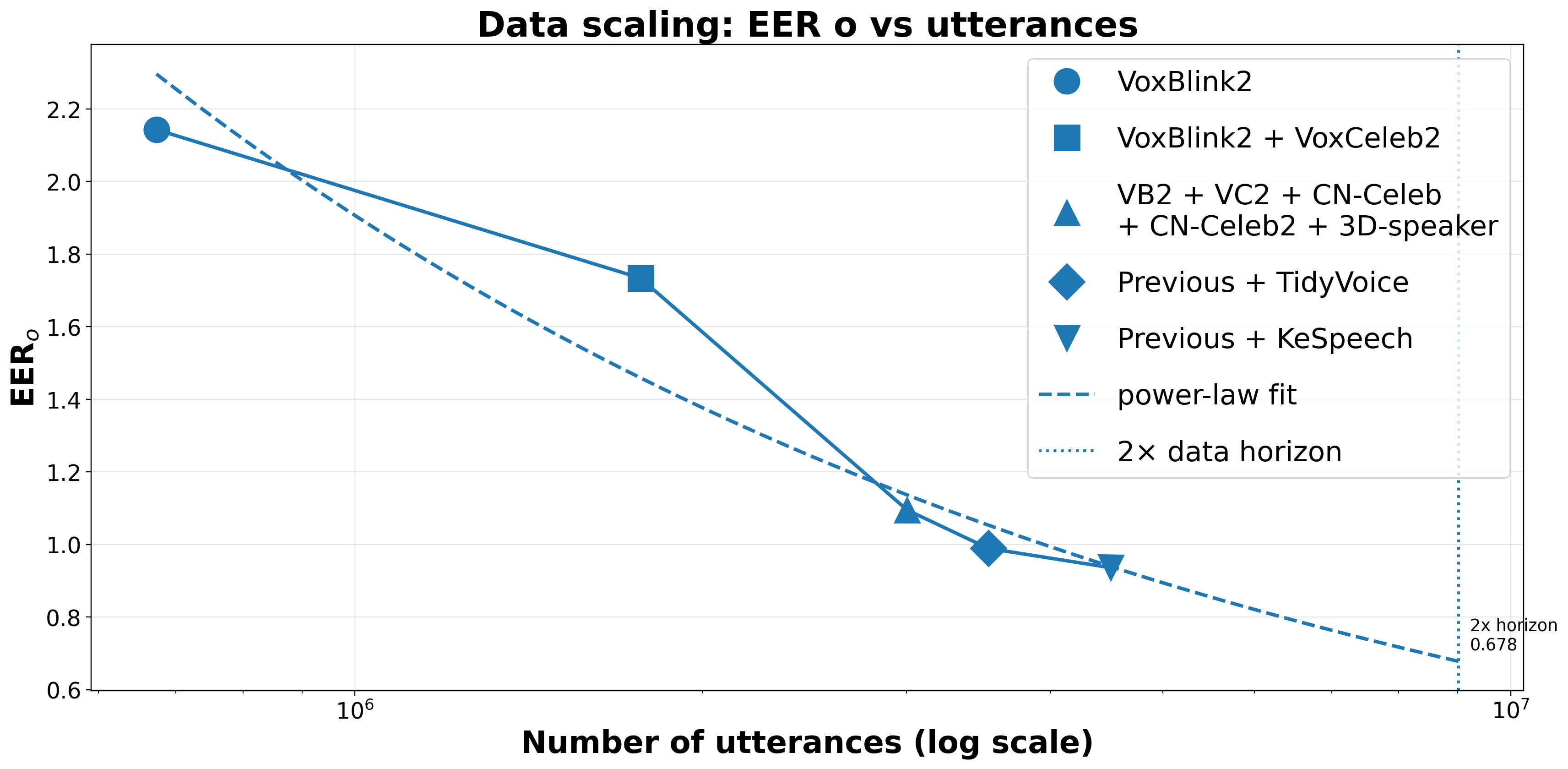}}
\hfil
\subfloat[]{\includegraphics[width=0.32\textwidth]{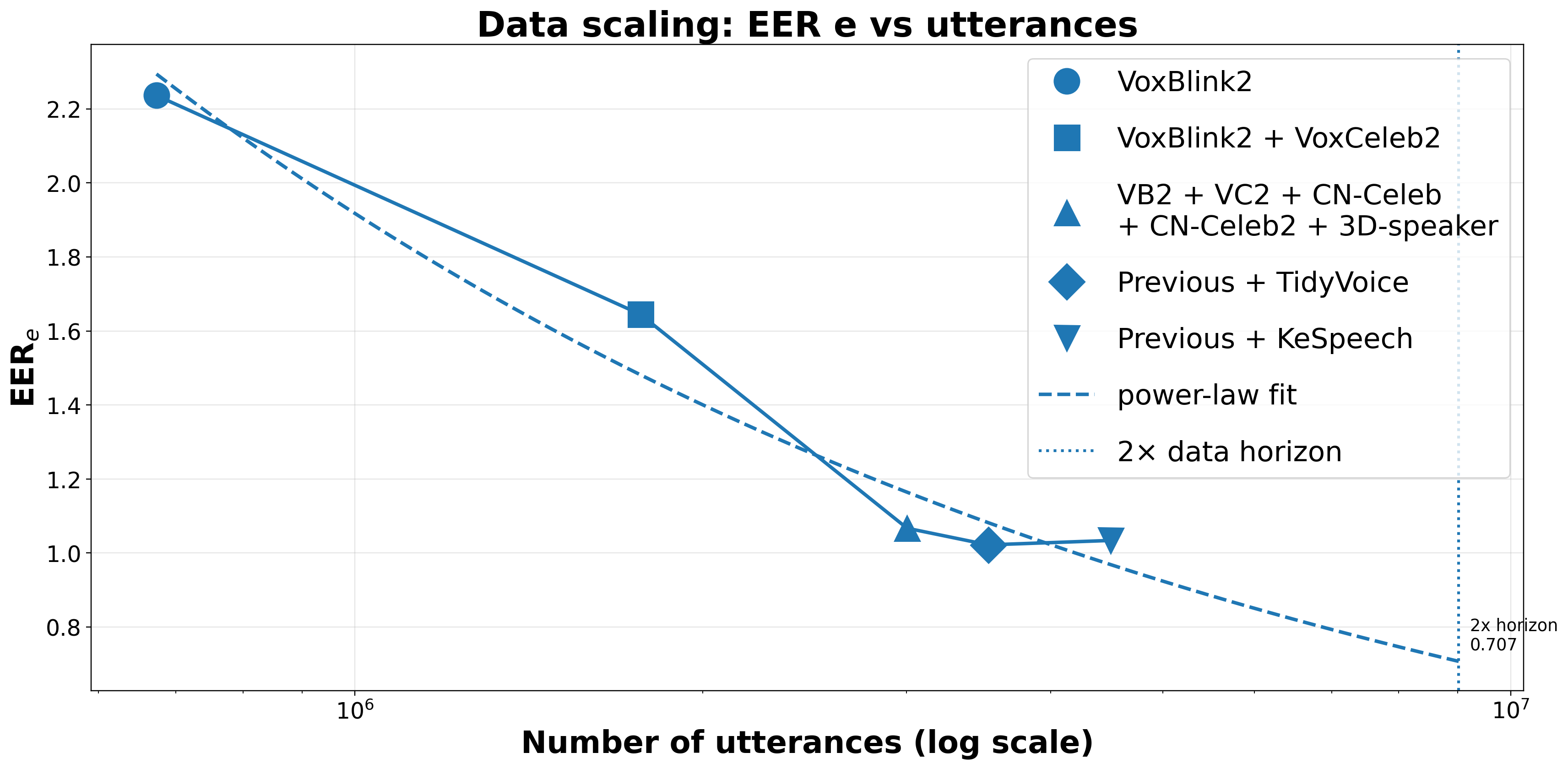}}
\hfil
\subfloat[]{\includegraphics[width=0.32\textwidth]{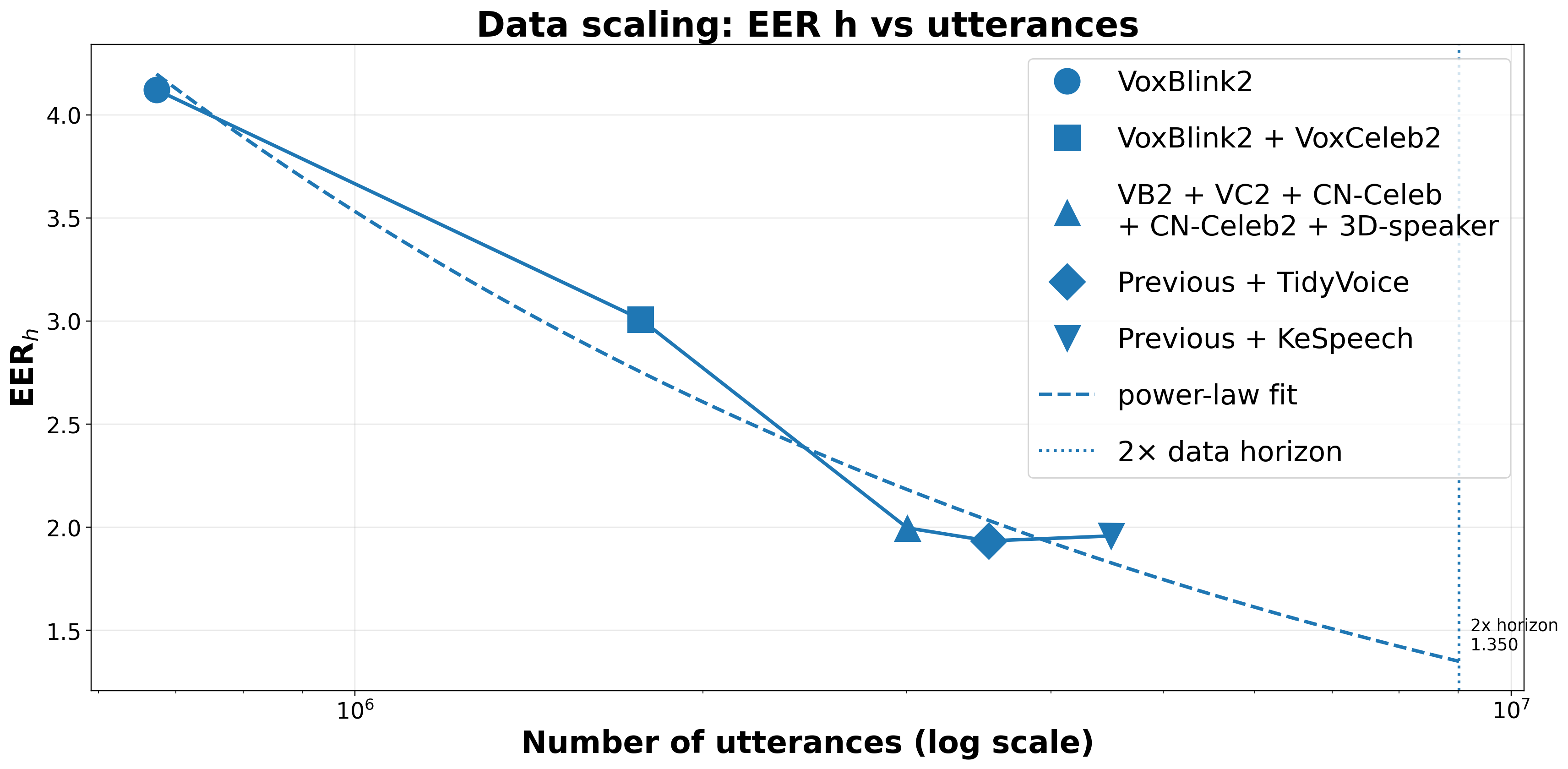}}
\caption{Data scaling trends for VoxCeleb1-O, VoxCeleb1-E, and VoxCeleb1-H as a function of training speakers and training utterances.}
\label{fig:scaling}
\end{figure*}

\subsection{External Baselines}
Table~\ref{tab:external_results} reports the local VoxCeleb1-O/E/H comparison with 4-second windows for every model. ReDimNet2+ LMFT reaches 0.351 / 0.523 / 1.055\% EER and is lower than all three reported WeSpeaker ONNX checkpoints on every protocol. Training budgets and corpora differ, so this comparison evaluates final systems under a shared local protocol rather than matched training data.

\begin{table*}[!t]
\caption{Local VoxCeleb1-O/E/H Comparison with 4-Second Windows\label{tab:external_results}}
\centering
\footnotesize
\begin{tabular}{llrrr}
\toprule
System & Runtime & EER$_o$ (\%) & EER$_e$ (\%) & EER$_h$ (\%) \\
\midrule
ReDimNet2+ LMFT & Torch & \textbf{0.351} & \textbf{0.523} & \textbf{1.055} \\
WeSpeaker CAM++ & ONNX/CUDA & 0.787 & 0.928 & 1.824 \\
WeSpeaker ResNet34-LM & ONNX/CUDA & 0.814 & 0.933 & 1.679 \\
WeSpeaker ECAPA512-LM & ONNX/CUDA & 0.877 & 1.071 & 1.968 \\
\bottomrule
\end{tabular}
\end{table*}

\subsection{Reranking Results}
Table~\ref{tab:rerank_results} shows the reranking ablation on the ReDimNet2+ pretrained setup: Pr@1/10/45 are measured on VoxCeleb1, while the VB2 column is measured on the VoxBlink2 subset. On VoxBlink2, mean-chain reranking raises Pr@$k$ from 0.7024 to 0.7220. Graph reranking alone reaches 0.7369. The combined mean-chain plus graph variant reaches 0.7391. Applied to the final LMFT system, the same reranking pipeline reaches 0.7687 Pr@$k$, compared with 0.6906 for the baseline ReDimNet2 after the same reranking.


\subsection{Inference}
Table~\ref{tab:inference} reports a separate inference-runtime comparison on fixed 6-second segments. TensorRT FP16 reaches 230.66$\times$ real time, with EER close to Torch FP32 in the same 6-second setup. Here xRTF is processed audio duration divided by elapsed processing time, rather than utterances per second. These accuracy figures are separate from the main 4-second evaluation.

\begin{table*}[!t]
\caption{Inference Speed and Quality on Fixed 6-Second Segments. xRTF is audio duration divided by processing time; higher is faster\label{tab:inference}}
\centering
\footnotesize
\begin{tabular}{lrrrrrr}
\toprule
Implementation & xRTF & Batch & Workers & EER$_o$ & EER$_p$ & EER$_{ph}$ \\
\midrule
Baseline ONNX FP32 & 707.61 & 32 & 4 & 14.647 & 17.197 & 41.634 \\
Ours Torch FP32 & 59.09 & 32 & 4 & 0.489 & 1.076 & 8.995 \\
Ours Torch BF16 & 82.53 & 32 & 4 & 0.484 & 1.072 & 8.968 \\
Ours ONNX FP32 & 61.87 & 32 & 4 & 0.489 & 1.075 & 9.000 \\
Ours ONNX FP16 & 113.66 & 32 & 4 & 0.489 & 1.076 & 8.984 \\
Ours TensorRT FP32 & 116.43 & 32 & 4 & 0.489 & 1.076 & 9.000 \\
Ours TensorRT BF16 & 117.04 & 32 & 4 & 0.489 & 1.075 & 8.968 \\
Ours TensorRT FP16 & \textbf{230.66} & 32 & 4 & 0.484 & 1.075 & 8.989 \\
\bottomrule
\end{tabular}
\end{table*}

\section{Discussion}
\label{app:discussion}
\subsection{Main Sources of Improvement}
The development sequence combines broader training data, longer training windows, and LMFT. Adding device, distance, genre, and language diversity accompanies improvements in verification and retrieval. Increasing the training window may provide more speaker evidence for longer recordings, while LMFT combines 6-second training segments with a stronger margin. The evaluation window remains 4 seconds. Because multiple factors change between runs, the sequence supports the combined recipe but does not isolate each intervention's causal effect.

\subsection{Motivation for Codec Augmentation}
The exploratory analysis reveals a shift in predicted coloration and FLAC compressibility. These observations motivate codec simulation and filtering alongside additive noise and reverberation. The transformations expose training to bandwidth changes and compression artifacts, but the observed shift does not prove that codec augmentation is necessary for every target domain. FLAC encoding itself is lossless and should not be interpreted as lossy-codec corruption.

\subsection{Verification Versus Retrieval}
The results also show that verification and retrieval need separate treatment. EER measures threshold behavior on labeled pairs, while Pr@$k$ measures ranking quality in a gallery. The graph reranking stage improves the latter by using structure that is invisible to pairwise scoring alone: reciprocal neighbors, shared neighbors, and hubness. This matters when many embeddings are close to each other and a few universal embeddings appear in many top-$k$ lists.

\subsection{Remaining Failure Modes}
The largest remaining gaps appear when the evaluation protocol concentrates difficult impostor pairs or applies explicit signal degradation. For the final model, EER$_h$ is about twice EER$_e$, indicating that the hard protocol still exposes confusions among speakers with more similar metadata or acoustic conditions. EER$_{ph}$ remains higher than the clean VoxCeleb1 metrics even after LMFT, which means that the codec and waveform stress suite still contains transformations that are not fully absorbed by the embedding space. These observations suggest that future work should focus on real telephone/VoIP data, more systematic codec curriculum design, and calibration under degraded-channel score distributions.

\subsection{Limitations}
This work has several limitations. First, the final ReDimNet2+ numbers are produced by a staged recipe with multiple corpora, so external open checkpoints are not controlled for training data; the local comparison is informative under a shared protocol but is not an equal-data architecture comparison. Second, EER$_{ph}$ uses synthetic degradations; it measures robustness to a chosen transformation set and need not cover all real-world channel failures. Third, the separate runtime benchmark uses 6-second inputs, so its accuracy figures should not be conflated with the main 4-second evaluation. Fourth, a major remaining challenge is out-of-domain generalization on difficult open-set targets; sub-1\% VoxCeleb1 EER alone is not sufficient evidence of robustness for arbitrary in-the-wild deployments.

\section{Reproducibility}
\label{app:repro}
\noindent Code: \href{https://github.com/lab260ru/redimnet2-plus}{github.com/lab260ru/redimnet2-plus}.\\
Weights: \href{https://huggingface.co/lab260/redimnet2-plus}{huggingface.co/lab260/redimnet2-plus}.

All experiments use a fixed seed of 42, BFloat16 mixed precision, AdamW with gradient clipping at 20, and distributed execution with Accelerate and FSDP. The recorded ReDimNet2+ artifacts include the model code, per-stage configuration files for the backbone, loss, augmentation, and data pipelines, and the run-level hyperparameters reported in Table~\ref{tab:hparams}.

For the external comparison, we record model and ONNX checkpoint checksums, the trial files, per-utterance embeddings, per-trial scores, aggregate metric tables, and an environment snapshot containing a pip freeze, GPU and driver information, and kernel metadata. The three reported WeSpeaker systems were benchmarked on an NVIDIA GeForce RTX 4080 SUPER under a single Python environment. The benchmark code is covered by nine unit tests for trial parsing, utterance collection, EER computation, score formation, audio-path resolution, PCM-scale audio loading, shard partitioning, O/E/H aggregation, and reuse of a superset embedding cache when scoring VoxCeleb1-H from VoxCeleb1-E embeddings.

\bibliographystyleapp{IEEEbib}
\bibliographyapp{references}
\fi

\end{document}